\documentclass[ amsmath,amssymb,onecolumn,longbibliography,
aps,floatfix]{revtex4-1}

\usepackage{graphicx}

\usepackage{dblfloatfix}

\usepackage{dcolumn}
\usepackage{bm}
\usepackage{amsmath}
\usepackage{amsfonts}
\usepackage[colorlinks=true, citecolor=blue, linkcolor=blue,urlcolor=blue ]{hyperref}

\usepackage{hyperref} 

\usepackage[dvipsnames]{xcolor}

\usepackage{siunitx}

\usepackage{listings}
\usepackage{color}
\usepackage{caption}
\usepackage{orcidlink}

\hypersetup
{
	pdfauthor={Weida Liao},
	pdftitle={Hydrodynamic mechanism for stable spindle positioning in meiosis II oocytes},
}

\AtBeginDocument{%
	\newwrite\bibnotes
	\def\bibnotesext{Notes.bib}
	\immediate\openout\bibnotes=\jobname\bibnotesext
	\immediate\write\bibnotes{@CONTROL{REVTEX41Control}}
	\immediate\write\bibnotes{@CONTROL{%
			apsrev41Control,author="08",editor="1",pages="1",title="0",year="1"}}
	\if@filesw
	\immediate\write\@auxout{\string\citation{apsrev41Control}}%
	\fi
}%

\begin{document}

	\begin{titlepage}

		\title{Hydrodynamic mechanism for stable spindle positioning in meiosis II oocytes}

		\date{\today}
		\author{Weida Liao\,\orcidlink{0000-0002-0000-228X}}
		\affiliation{Department of Applied Mathematics and Theoretical Physics, University of Cambridge, Wilberforce Road,  Cambridge, CB3 0WA, UK}
		\author{Eric Lauga\,\orcidlink{0000-0002-8916-2545}}
		\email{e.lauga@damtp.cam.ac.uk}
		\affiliation{Department of Applied Mathematics and Theoretical Physics, University of Cambridge, Wilberforce Road,  Cambridge, CB3 0WA, UK}
		\begin{abstract} 
Cytoplasmic streaming, the persistent flow of fluid inside a cell, induces intracellular transport, which plays a key role in fundamental biological processes. In meiosis II mouse oocytes (developing egg cells) awaiting fertilisation, the spindle, which is the protein structure responsible for dividing genetic material in a cell, must maintain its position near the cell cortex (the thin actin network bound to the cell membrane) for many hours. However, the cytoplasmic streaming that accompanies this stable positioning would intuitively appear to destabilise the spindle position. Here, through a combination of numerical and analytical modelling, we reveal a new, hydrodynamic mechanism for stable spindle positioning beneath the cortical cap. We show that this stability depends critically on the spindle size and the active driving from the cortex, and demonstrate that stable spindle positioning can result purely from a hydrodynamic suction force exerted on the spindle by the cytoplasmic flow. Our findings show that local fluid dynamic forces can be sufficient to stabilise the spindle, explaining robustness against perturbations not only perpendicular but also parallel to the cortex. Our results shed light on the importance of cytoplasmic streaming in mammalian meiosis.

		\end{abstract}
		
		\maketitle

	\end{titlepage}
	
	\newpage

\section{Introduction}\label{sec:introduction}

\subsection{Transport by fluid flow in biology}

From ocean currents carrying plankton to blood circulation in the cardiovascular system, transport by fluid flow plays a key role in many diverse biological processes, spanning a wide range of length scales~\cite{cartwright2009fluid}.
From a biological standpoint, this rich spectrum covers systems from the biosphere and communities, to organisms and their fundamental building blocks: individual cells. 
One of the defining characteristics of living organisms is the capacity for reproduction. 
Many striking examples of transport by fluid flow can be drawn from reproductive and developmental biology at the microscopic scale, where flow is dominated by viscous forces.
For instance, in developing embryos, unidirectional flow driven by hair-like appendages on the surface of cells (cilia) establish left--right asymmetry of the organism~\cite{nonaka2005novo}. 
Even earlier in development, and at the level of just one cell, the inertialess flow of the fluid inside an egg cell during cell division influences its subsequent development, by inducing transport and hence positioning of important structures within the cell~\cite{yi2011dynamic}.  
In our work, we reveal a physical mechanism by which this positioning can be achieved robustly against perturbations. 

\subsection{Meiosis}

To provide the relevant biological background, we begin in this section with a brief introduction to meiosis, the specialised cell division that generates reproductive cells; in the next section, we will zoom in on the biophysical problem at the heart of our study. 
The egg cell is the female reproductive cell, which develops from a precursor cell called an oocyte. 
For the mature egg cell to contain the correct amount of genetic material, the oocyte must undergo two specialised divisions, known as meiosis~I and meiosis~II~\cite{alberts2002molecular}.
Notably, the two meiotic divisions are highly asymmetric~\cite{mogessie2018assembly}, each removing excess chromosomes from the large oocyte into a much smaller cell, a by-product of the cell division that then typically degenerates~\cite{schmerler2011polar,li2013road}.  
The egg cell is filled with a complex fluid called cytoplasm~\cite{mogilner2018intracellular,luby1999cytoarchitecture}, which contains important resources (e.g.~proteins and cellular organelles) required for embryo development. 
For the mature egg cell to contain as much cellular material as possible, it is therefore crucial for the oocyte to minimise the amount of cytoplasm removed during the two meiotic divisions, while still eliminating the surplus chromosomes~\cite{brunet2011positioning,mogessie2018assembly,dalton2013biased}.

\begin{figure}[t]
	{\includegraphics[width=0.85\textwidth]{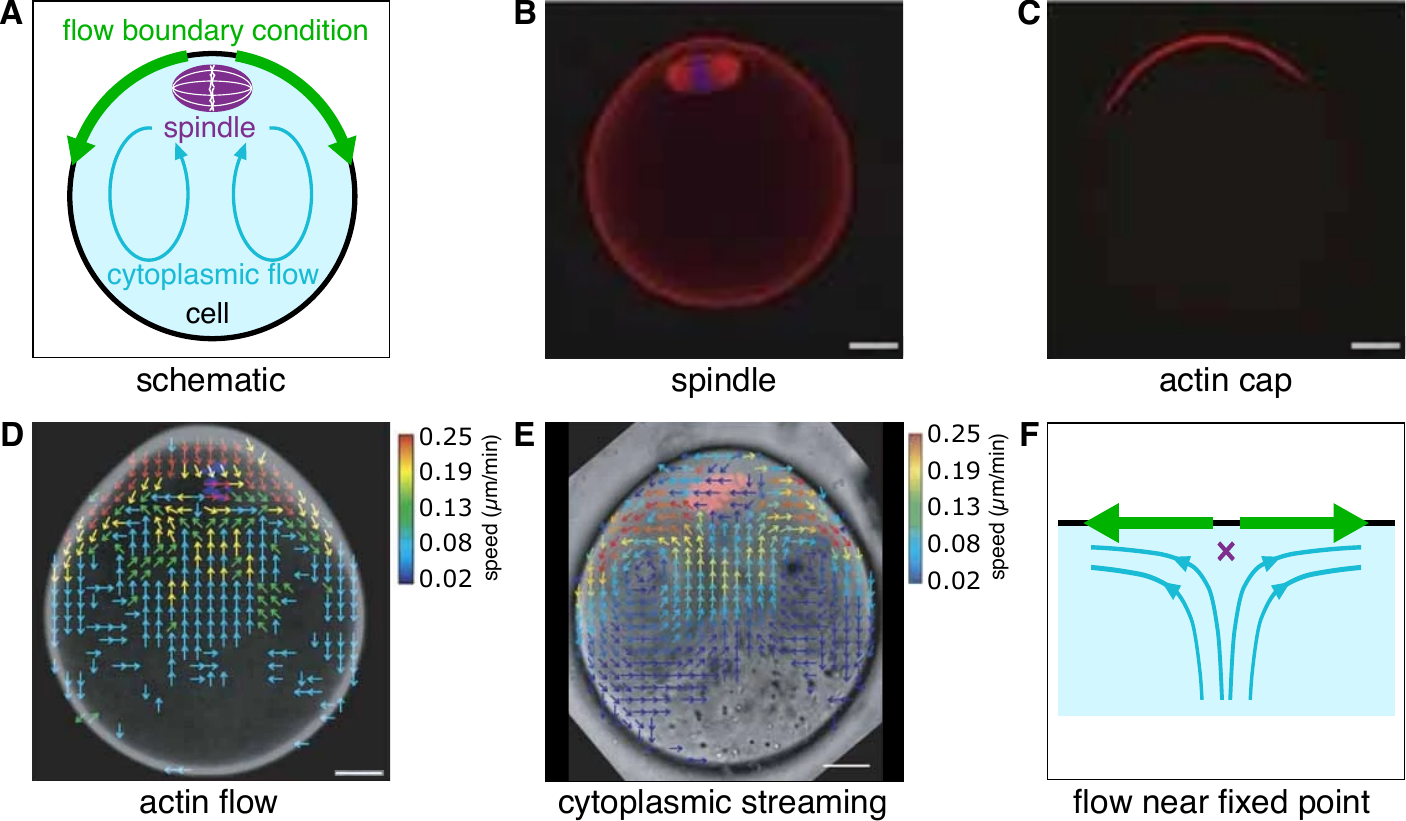}
		\caption{
			Mouse oocyte in meiosis II. 
				(A)~Schematic of model for spindle positioning in meiosis II oocyte; 
a localised flow boundary condition (green) drives cytoplasmic streaming (blue) in the bulk of the cell, keeping the spindle (purple) near the cortex. 
				(B)~Spindle in oocyte, in which  DNA (chromatin) is shown in blue, while microtubules are shown in red  
				(scale bar, $10~\si{\micro\metre}$).
				(C)~Actin-rich cortical cap (red) of the oocyte (scale bar, $10~\si{\micro\metre}$).
				(D)~Actin flow field inside the cell (scale bar, $10~\si{\micro\metre}$). 
				(E)~Cytoplasmic flow field, with 
				DNA (chromatin) shown in pink (scale bar, $10~\si{\micro\metre}$).
				(F)~Schematic of local flow (blue) close to the fixed point (purple), the position at which the spindle remains, near the cell cortex, where the active slip velocity boundary condition (green) is applied. 
				Panels B--E adapted  from Ref.~\cite{yi2011dynamic} and reproduced with permission. 
				} 
		\label{fig:diagram_motivating_spindle_stability}}
\end{figure}

A key piece of apparatus involved in meiosis is the meiotic spindle, which we illustrate schematically in Fig.~\ref{fig:diagram_motivating_spindle_stability}A (purple). 
The spindle is the protein structure responsible for dividing the genetic material in the cell, and it assembles in each of the two meiotic divisions. 
The spindle includes the chromosomes, attached to long polymers called microtubules. 
Importantly, its asymmetric positioning in the oocyte is responsible for the asymmetric division of the cell~\cite{almonacid2014actin,li2013art}; indeed, loss of asymmetric spindle positioning is seen in ageing oocytes~\cite{chaigne2012spindle,webb1986parthenogenesis,kim1996cytoskeletal,yi2012actin}, which are associated with reduced development potential~\cite{miao2009oocyte}. 
To achieve the required asymmetric cell division, the spindle must be positioned not at the centre of the oocyte, but instead close to the cell cortex, a thin, dense network of proteins bound to the cell membrane (i.e.~at the boundary of the cell)~\cite{chugh2018actin}.
Through experiments and biophysical modelling, various studies have explored how asymmetric spindle positioning is established in meiosis I for mammalian oocytes~\cite{yi2013sequential,yi2013symmetry,duan2020dynamic,schuh2008new,chaigne2013soft,duan2019actin,xie2018poly,maddox2012polar,chaigne2015narrow}.
After meiosis I, the second meiotic spindle rapidly forms around the chromosomes positioned near the cortex~\cite{azoury2008spindle,yi2011dynamic}.

\subsection{Stable spindle positioning by cytoplasmic streaming in meiosis II oocytes}

Before the oocyte divides at the end of meiosis II, the spindle remains stably positioned near the cortex for many hours, as the oocyte awaits fertilisation. 
The aim of our work is to rationalise the stable positioning of the spindle in this stage of meiosis II through biophysical modelling. 
Experiments have revealed intracellular flow during this stage of meiosis II~\cite{yi2011dynamic,niwayama2016bayesian}.
This is an example of cytoplasmic streaming: actively-driven, persistent, bulk flow of the cytoplasm inside a cell~\cite{lu2023go,shamipour2021cytoplasm,mogilner2018intracellular}.
Cytoplasmic streaming occurs inside cells of many different organisms, such as  slime moulds~\cite{alim2013random,kamiya1981physical}, fungi, algae and higher plants~\cite{goldstein2015physical,allen1978cytoplasmic}.  
Several examples of cytoplasmic streaming have also been characterised in the largest animal cells~\cite{quinlan2016cytoplasmic,klughammer2018cytoplasmic,gubieda2020going}, including oocytes of various species during different developmental stages.
Multiple mechanisms, involving various types of polymeric filaments and molecular motors,  are able to generate cytoplasmic motion~\cite{lu2023go}. 

In mouse oocytes near completion of meiosis II, cytoplasmic streaming is driven by the flow of actin (a polymeric filament) away from the cortical actin cap~\cite{yi2011dynamic} (Fig.~\ref{fig:diagram_motivating_spindle_stability}). 
Based on a combination of experiments and fluid dynamical simulations, the authors of Ref.~\cite{yi2011dynamic} postulated that the cytoplasmic streaming would push the spindle towards the cortex, thus stably maintaining the asymmetric spindle position near the cortex. 
However, the diverging cytoplasmic flow along the cortex appears intuitively to destabilise the spindle position, and the physical mechanism for the stability observed in spite of this remains to be elucidated. 
In this article, we demonstrate through hydrodynamic modelling that the cytoplasmic flow can exert a suction (pulling force) on the spindle towards the cortex.
This in turn allows stable positioning of the spindle near the cortical cap, with robustness against perturbations not only perpendicular but also parallel to the cortex. 

We illustrate in Fig.~\ref{fig:diagram_motivating_spindle_stability} the biological context of our paper, using mouse oocytes in meiosis II; 
in panels B--E, we have adapted figures from Ref.~\cite{yi2011dynamic}. 
The mouse oocyte is a model system for understanding mechanisms for asymmetric spindle positioning in mammals~\cite{mogessie2018assembly,uraji2018functions}. 
Our modelling approach is shown schematically in Fig.~\ref{fig:diagram_motivating_spindle_stability}A, where an active slip velocity (green) at the cell cortex drives the bulk intracellular fluid flow (blue). 
The meiotic spindle is positioned inside the oocyte near the cortical cap, at the top of the experimental image in Fig.~\ref{fig:diagram_motivating_spindle_stability}B; DNA is shown in blue and microtubules in red. 
The cortical cap of the oocyte, in Fig.~\ref{fig:diagram_motivating_spindle_stability}C,  is rich in actin (red). 
Actin filaments are nucleated by a protein complex localised to the cortical cap, and flow continuously away from it~\cite{yi2011dynamic}. 
The actin flow in the oocyte, quantified in Fig.~\ref{fig:diagram_motivating_spindle_stability}D, has the highest velocity as it leaves the cortical cap  {(at the top of the image)}. 
We observe a toroidal actin flow field, downwards along the cell periphery, and recirculating upwards towards the actin cap through the centre of the oocyte~\cite{yi2011dynamic,bourdais2021cofilin}.  
The actin flow depends on actin polymerisation and turnover~\cite{yi2011dynamic,pinot2012confinement,cramer1997molecular}.
This is consistent with a process known as treadmilling, where the actin filaments elongate at one end and depolymerise from the other end~\cite{yi2012actin,lappalainen2022biochemical}.
Past biophysical modelling has revealed a mechanism by which polymerising actin filaments can generate force inside cells~\cite{mogilner1996cell}. 

The actin flow drives cytoplasmic streaming in the oocytes~\cite{yi2011dynamic}, shown in Fig.~\ref{fig:diagram_motivating_spindle_stability}E.
The cytoplasmic streaming follows a similar pattern to the actin flow: the small cytoplasmic particles are transported away from the cortical cap, downwards along the cell periphery and back upwards through the centre of the oocyte~\cite{yi2011dynamic,niwayama2016bayesian}. 
Importantly, the cytoplasmic flow has highest velocity near the cortical cap, reflected by the slip velocity at the cortex that drives the flow in our model (Fig.~\ref{fig:diagram_motivating_spindle_stability}A). 

The nature of the cytoplasmic flow close to the cortical cap raises an intriguing question on the stability of spindle positioning against perturbations along the cortex. 
Locally, the cytoplasmic flow near the spindle is similar to an extensional flow, as depicted schematically in Fig.~\ref{fig:diagram_motivating_spindle_stability}F (blue).
The spindle remains stably for many hours at the fixed point of the flow (purple), as the oocyte waits for fertilisation. 
However, this fixed point appears intuitively to be unstable, when considering perturbations parallel to the cortex: if a small tracer is placed just to the right of the fixed point, then it is advected rightwards by the fluid flow, away from the fixed point (in a linear stagnation point flow, this would in fact happen  exponentially fast~\cite{guyon2015physical}).

The goal of our work is to explain physically the stable positioning of the spindle at the fixed point. 
We demonstrate through physical modelling that the cytoplasmic flow can create a hydrodynamic (suction) force on the spindle towards the fixed point.
We thus show that fluid dynamic forces can be sufficient for stable spindle positioning, depending critically on the size of the spindle and the flow forcing due to the cortical cap. 

\subsection{Structure of paper}

This article is organised as follows. 
First, in Sec.~\ref{sec:flow_simulation}, we introduce our model for the cytoplasmic flow inside an oocyte, actively driven by a slip velocity at the cortex and based on experimental measurements of cytoplasmic streaming. 
We next present in Sec.~\ref{sec:forces_spherical} and Sec.~\ref{sec:phase_diagram_spherical} our model for the spindle inside the oocyte. 
We numerically compute the force on the spindle due to the flow and find that the stability of the fixed point depends on two key parameters: the spindle radius and the size of the active slip domain at the cortex. 
In Sec.~\ref{sec:force_simple} and Sec.~\ref{sec:phase_diagram_simple}, we explain these stability results by introducing an intuitive physical model for the spindle in flow driven by a slip velocity. 
Solving analytically for the flow, we obtain the force on the spindle and hence the stability of the fixed point as a function of the spindle radius and slip domain size within this model. 
Importantly, our simple analytical model reproduces the stability trends from our numerical simulations. 
This allows us to elucidate the physical mechanism for the stable spindle positioning, in terms of the competing effects of the shear and suction due to the fluid flow, in Sec.~\ref{sec:physical_mechanism}. 
In Sec.~\ref{sec:discussion}, we summarise our results and discuss them in the context of experimental evidence on spindle positioning. 
We close by outlining possible future experiments and modelling ideas. 

\section{Results}

\subsection{Simulated cortex-driven cytoplasmic streaming matches experimental flow field}\label{sec:flow_simulation}

\begin{figure}[t]
	{\includegraphics[width=0.75\textwidth]{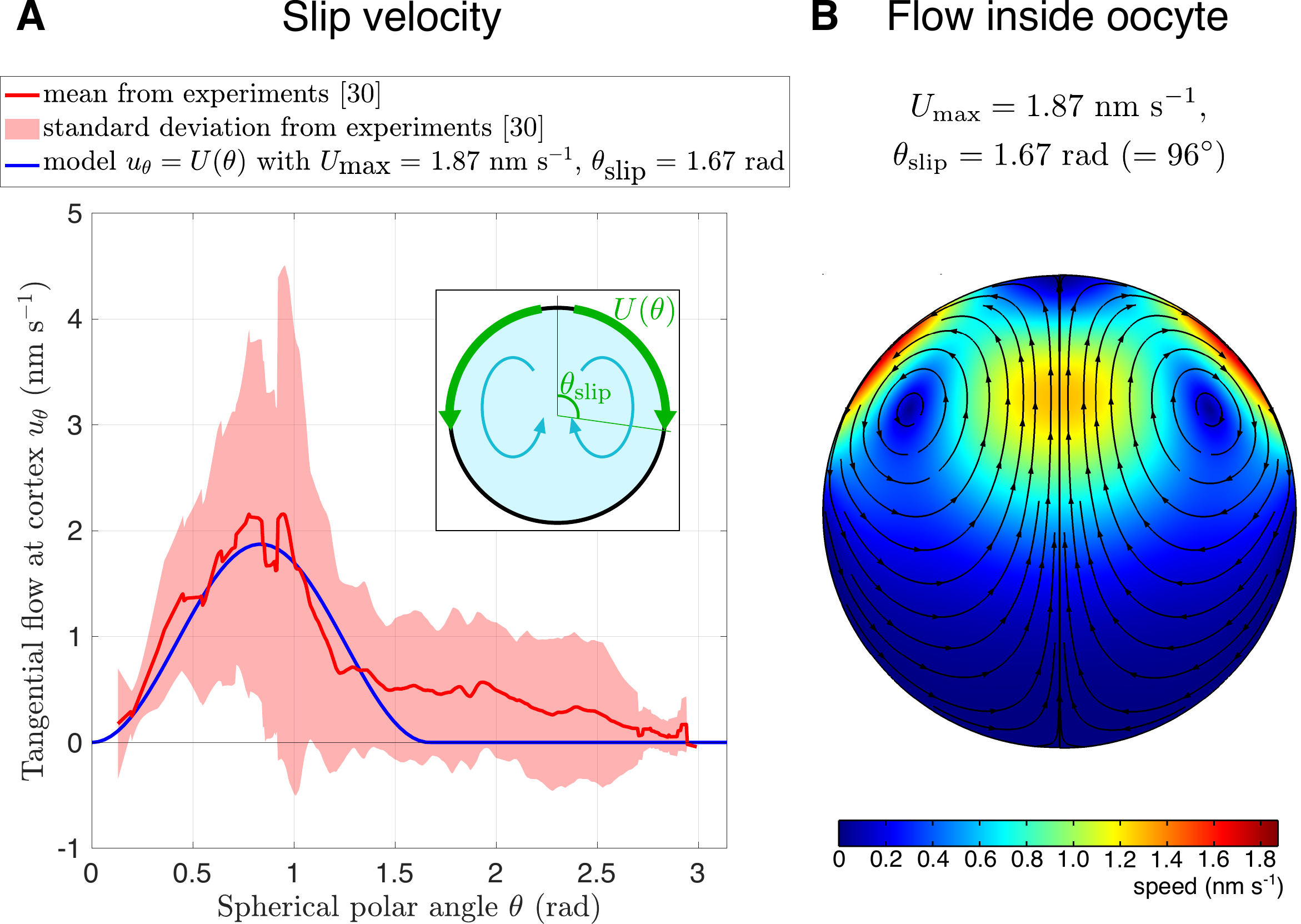} 
		\caption{
(A)~Tangential flow $u_\theta$ at the cortex of mouse oocytes as a function of spherical polar angle $\theta$, from experimental data~\cite{niwayama2016bayesian} (red) and for model fitted to data (blue), together with model setup (inset); red curve: experimental mean; light red shading: standard deviation; 
blue curve: model, $u_\theta=U(\theta)$ from Eq.~\eqref{eq:U_theta_slip_velocity}, with $U_\text{max}=1.87~\si{\nano\metre}~\si{\second}^{-1}$,~$\theta_\text{slip}=1.67~\si{\radian}~$($=\ang{96}$) from least squares fit. 
			Inset: a prescribed active slip velocity $u_\theta=U(\theta)$ (green), which is nonzero inside a spherical polar angle $\theta\leq\theta_\text{slip}$ in our model, drives a cytoplasmic flow (blue arrows) inside the cell.
			(B)~Flow inside a spherical model oocyte as obtained by numerical simulation, induced by a prescribed active slip velocity boundary condition at cortex, with model parameters taken from least squares fit. 
			}
		\label{fig:plot_Niwayama_2016_least_squares_nlinfit_plot_v4_27022024_with_inset_v1_27022024}}
\end{figure}

Here we introduce our model for the cytoplasmic flow inside the oocyte, based on experimental measurements~\cite{yi2011dynamic,niwayama2016bayesian}. 
This enables us to identify the parameters that characterise the forcing from the actin cap. 
We will then build on this in Sec.~\ref{sec:forces_spherical}, where we add the spindle to the geometry and examine the effect of cortex-driven flow on it.

We model the cytoplasmic streaming as incompressible Stokes flow inside a sphere, driven by a prescribed active slip velocity boundary condition~(Fig.~\ref{fig:plot_Niwayama_2016_least_squares_nlinfit_plot_v4_27022024_with_inset_v1_27022024} and see Sec.~\ref{sec:flow_model} for details). 
This is motivated by experimental measurements of the cytoplasmic flow field in meiosis II mouse oocytes from Refs.~\cite{yi2011dynamic} (Fig.~\ref{fig:diagram_motivating_spindle_stability}E) and~\cite{niwayama2016bayesian}, which is captured well by Stokes flow driven at the cortex~\cite{niwayama2016bayesian} (see Sec.~\ref{sec:discussion} for further discussion of model assumptions). 

In Fig.~\ref{fig:plot_Niwayama_2016_least_squares_nlinfit_plot_v4_27022024_with_inset_v1_27022024}A, we plot the tangential flow velocity $u_\theta$ at the cortex from experiments~\cite{niwayama2016bayesian}, averaged over the oocytes, as a function of the spherical polar angle $\theta$ (red curve), and we indicate the standard deviation (light red shading). 
The blue curve illustrates our model for the tangential flow at the cortex (setup in inset of Fig.~\ref{fig:plot_Niwayama_2016_least_squares_nlinfit_plot_v4_27022024_with_inset_v1_27022024}A) fitted to the experimental data. 
The active slip velocity $u_\theta=U(\theta)$ [Eq.~\eqref{eq:U_theta_slip_velocity}] models the forcing due to the actin flow from the cortical cap. 
It is parametrised only by the maximum velocity $U_\text{max}$ and the slip angle $\theta_\text{slip}$ (the size of the active slip domain where the slip velocity is positive, green in Fig.~\ref{fig:plot_Niwayama_2016_least_squares_nlinfit_plot_v4_27022024_with_inset_v1_27022024}A inset). 
Its highly localised nature is based on the experimental flow field in Fig.~\ref{fig:diagram_motivating_spindle_stability}E, reproduced from Ref.~\cite{yi2011dynamic}, the experiments from which our study aims to explain. 
In order to estimate biological parameter values, we will employ the data from Ref.~\cite{niwayama2016bayesian}, obtained from several oocytes. 
The idealised form [Eq.~\eqref{eq:U_theta_slip_velocity}] approximates the experimental data, while allowing us to make analytical progress using our fundamental physical model in Sec.~\ref{sec:phase_diagram_simple}. 

In our later analysis of the stability of spindle positioning through numerical simulations (Sec.~\ref{sec:phase_diagram_spherical}), we vary the value of the slip angle $\theta_\text{slip}$, to explore the parameter space. 
However, for comparison with experiments~\cite{yi2011dynamic}, we use nonlinear least squares to fit the prescribed velocity in Eq.~\eqref{eq:U_theta_slip_velocity} to the experimental data~\cite{niwayama2016bayesian}. 
We thus obtain biological values for the relevant velocity scale and active slip angle in our model as $U_\text{max}=1.87~\si{\nano\metre}~\si{\second}^{-1}$,~$\theta_\text{slip}=1.67~\si{\radian}~$($=\ang{96}$), respectively, which we use for the model slip velocity $u_\theta=U(\theta)$ in blue in Fig.~\ref{fig:plot_Niwayama_2016_least_squares_nlinfit_plot_v4_27022024_with_inset_v1_27022024}A. 
We note that the active slip velocity therefore extends beyond the actin cap~(Fig.~\ref{fig:diagram_motivating_spindle_stability}C)~\cite{yi2011dynamic,niwayama2016bayesian}. 
To illustrate our model for the cytoplasmic streaming, we plot in Fig.~\ref{fig:plot_Niwayama_2016_least_squares_nlinfit_plot_v4_27022024_with_inset_v1_27022024}B the flow streamlines from numerical simulation, corresponding to these biological values. 
The resulting flow matches the toroidal flow field observed in experiments~\cite{yi2011dynamic}~(Fig.~\ref{fig:diagram_motivating_spindle_stability}E). 

\subsection{Hydrodynamic forces on spindle can lead to stable spindle positioning at fixed point}\label{sec:forces_spherical}

In Sec.~\ref{sec:flow_simulation}, we highlighted the active slip angle $\theta_\text{slip}$ as a key parameter in our model that characterises the cortex-driven cytoplasmic flow in the oocyte. 
To determine the conditions for stable spindle positioning by hydrodynamic forces, we now introduce the model spindle into the oocyte and demonstrate that our model contains sufficient physical ingredients to reproduce the stability seen experimentally. 
We will then analyse the robustness of this phenomenon in Sec.~\ref{sec:phase_diagram_spherical}, where we conduct a parameter sweep in our control parameters.

\begin{figure}[t]
	{\includegraphics[width=0.35\textwidth]{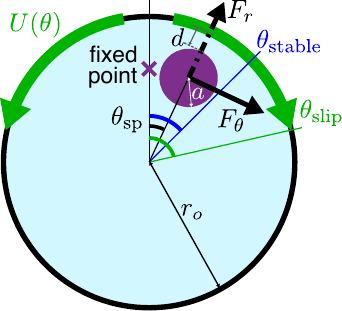}
		\caption{Model for spindle dynamics inside oocyte.
			The spindle (purple) is modelled as a sphere of radius~$a$ inside the oocyte of radius~$r_\text{o}$. 
			The spindle has angular position~$\theta=\theta_\text{sp}$.
			The clearance between the spindle and  cortex is denoted by $d$, a constant.
			The fixed point at $\theta=0$ is marked with a cross.
			The active slip velocity (green) is given by $u_\theta=U(\theta)$ from Eq.~\eqref{eq:U_theta_slip_velocity} and is nonzero in the angular domain $\theta \leq\theta_\text{slip}$, actively driving a fluid flow inside the cell. 
			The resulting hydrodynamic force exerted on the spindle has $r$ component $F_r$ (dashed black arrow) and $\theta$ component $F_\theta$ (solid black arrow). 
			Our simulations reveal a neighbourhood around the fixed point within which the spindle is hydrodynamically stable; the stability region for the fixed point is given by $0 \leq \theta_\text{sp} \leq \theta_\text{stable}$, indicated in blue. }
		\label{fig:fig_diagram_setup_v2_09112023}}
\end{figure}

We illustrate our model for the spindle dynamics in Fig.~\ref{fig:fig_diagram_setup_v2_09112023} (details in Sec.~\ref{sec:spindle_model}).
We model the spindle as a rigid sphere of radius $a$, and assume, based on experiments~\cite{yi2011dynamic}, that it is located close to the cortex, with constant clearance  $d=0.1r_\text{o}$ held fixed in our simulations.
This is a simplification of the geometry (Fig.~\ref{fig:diagram_motivating_spindle_stability}B) and the physics; however, the goal of our paper is to identify the fundamental physical ingredients required to understand the stable spindle positioning seen in experiments. 
In each numerical simulation, the spherical spindle is held stationary at position $\theta=\theta_\text{sp}$, with a no-slip boundary condition prescribed on its surface. 
We solve for the Stokes flow driven by the same active slip velocity as in Sec.~\ref{sec:flow_simulation} [Eq.~\eqref{eq:U_theta_slip_velocity}], but now in the annular region (fluid shaded in light blue in Fig.~\ref{fig:fig_diagram_setup_v2_09112023}) between the spindle and cell cortex.  
This cytoplasmic flow, modified in comparison with Fig.~\ref{fig:plot_Niwayama_2016_least_squares_nlinfit_plot_v4_27022024_with_inset_v1_27022024}B due to the addition of the spindle to the geometry, exerts hydrodynamic force on the fixed spindle, with radial component $F_r$ and $\theta$ component $F_\theta$.
We will show that the spindle can be hydrodynamically stable within a neighbourhood of the fixed point at $\theta=0$; this stability region for the fixed point has size~$\theta_\text{stable}$.

We can characterise the geometry and forces in the problem in terms of dimensionless ratios, i.e.~normalised physical quantities. 
First, we measure lengths in units of the oocyte radius $r_\text{o}$. 
We thus introduce the dimensionless spindle radius $\hat{a} \equiv a/r_\text{o}$ as the ratio of the spindle radius to the oocyte radius; similarly, the dimensionless oocyte radius is simply equal to $1$.

Next, we define dimensionless force components $\hat{F}_r\equiv F_r / 6\pi\mu a U_\text{max}$ and $\hat{F}_\theta\equiv F_\theta/ 6\pi\mu a U_\text{max}$, given by the ratio of the dimensional force to a characteristic force scale (the classical Stokes drag on a sphere~\cite{kimbook}), where $\mu$ is the dynamic viscosity of the fluid. 
For example, with dimensional cytoplasmic viscosity $\mu\approx 10^{2}~\si{\pascal}~\si{\second}$ (as measured for late meiosis I mouse oocytes)~\cite{chaigne2015narrow}, spindle radius $a\approx6~\si{\micro\metre}$~\cite{yi2011dynamic} and $U_\text{max}\approx2~\si{\nano\metre}~\si{s}^{-1}$ (Sec.~\ref{sec:flow_simulation}), a value of $\hat{F}_r=1$ for the dimensionless force on the spindle corresponds to a dimensional force $F_r\approx20~\si{\pico\newton}$.
This force scale is much larger than the typical root mean square force from thermal noise (around $0.004~\si{\pico\newton}$, by a scaling argument). 

In what follows, in order to simplify notation, we use the original variable names ($a$, $F_r$, $F_\theta$) to mean their dimensionless counterparts ($\hat{a}$, $\hat{F}_r$, $\hat{F}_\theta$, respectively) defined above. 
The control parameters of our model are therefore the dimensionless spindle radius $a$ (i.e.~the ratio of the spindle radius to the oocyte radius) and the size of the active slip domain $\theta_\text{slip}$. 

\begin{figure}[t] 
	{\includegraphics[width=0.75\textwidth]{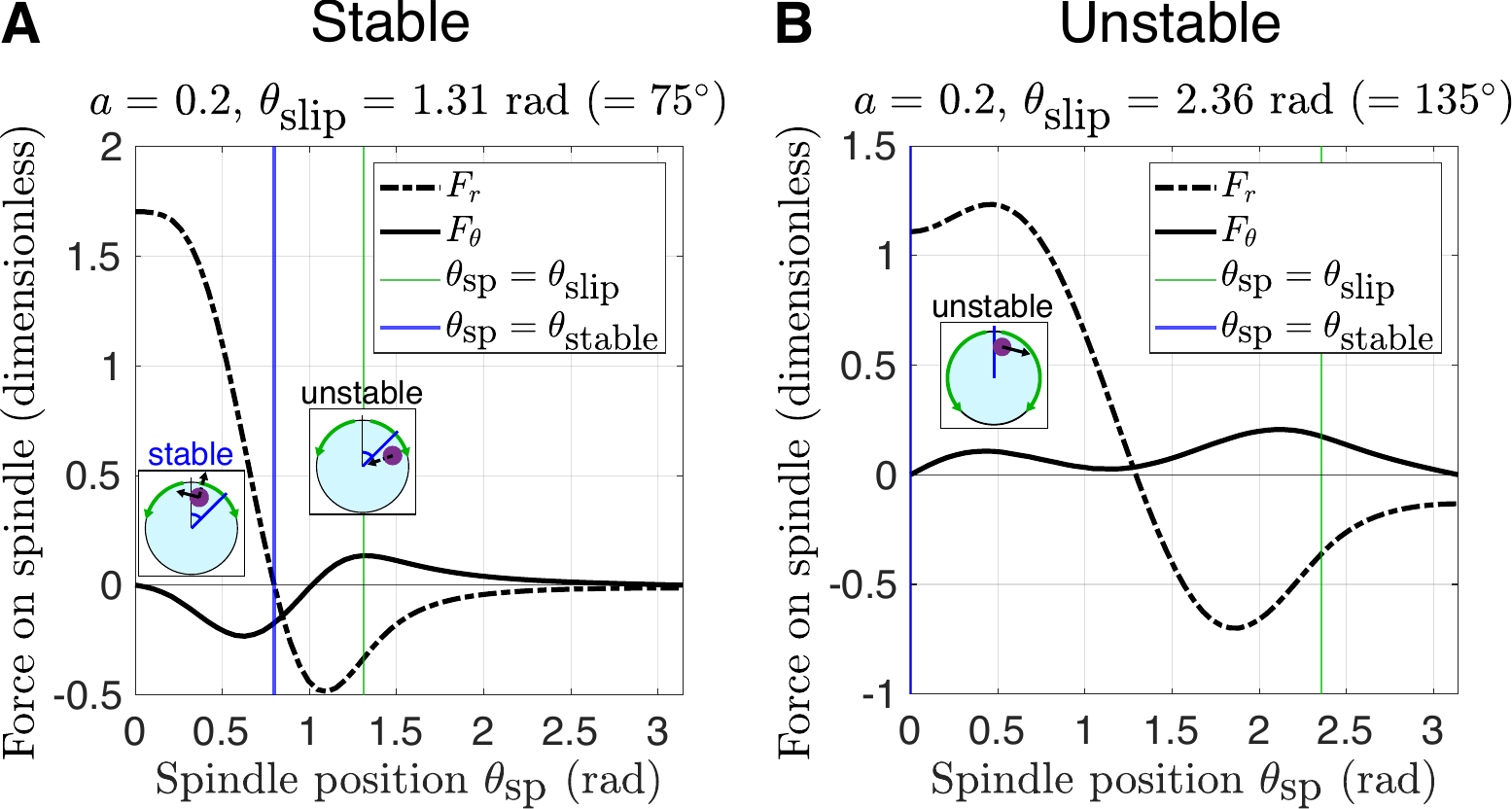} 
		\caption{
			Hydrodynamic forces on spindle [$F_r$ (dashed black) and $F_\theta$ (solid black)] vs angular position of spindle $\theta_\text{sp}$ (force normalisation explained in main text).
			(A)~Stable spindle positioning at fixed point $\theta_\text{sp}=0$ obtained for dimensionless spindle radius $a=0.2$ and slip angle $\theta_\text{slip}=1.31~\si{\radian}$~($=\ang{75}$).
			``Stable" inset: in the stability region $0\leq\theta_\text{sp}\leq \theta_\text{stable}$, the force satisfies $F_r \geq 0$ and $F_\theta \leq 0$; the fixed point is stable. 
			``Unstable" inset: $F_r<0$ outside the stability region (i.e.~$\theta_\text{sp} > \theta_\text{stable}$).
			(B)~Unstable spindle positioning at fixed point $\theta_\text{sp}=0$ for  $a=0.2$ and $\theta_\text{slip}=2.36~\si{\radian}$~($=\ang{135}$).  
			``Unstable" inset: $F_\theta >  0$.
			In both panels, the size of the stability region $\theta_\text{stable}$ is shown in blue, while the active slip angle $\theta_\text{slip}$ is in green. 
		}
		\label{fig:plot_41_45_force_fixed_sphere_closed_v7_stability_plot_forces_06022024_combined_v1_27022024}}
\end{figure}

To analyse the effect of the cortex-driven flow on the spindle, we compute the hydrodynamic force exerted on the fixed spindle as a function of position $\theta_\text{sp}$. 
Intuitively, we expect $\theta_\text{sp}=0$ to be a fixed point of the position of the spindle: the geometry is  axisymmetric in this case, while the flow in Fig.~\ref{fig:plot_Niwayama_2016_least_squares_nlinfit_plot_v4_27022024_with_inset_v1_27022024}B is upwards on the axis of symmetry. 
With our aim of understanding the stability of this fixed point, we can think of simply perturbing the position $\theta_\text{sp}$ of the spindle, and investigating: will the spindle return to the fixed point? 
If yes, how large is the stability region as a function of the key parameters of our model?  

We plot in Fig.~\ref{fig:plot_41_45_force_fixed_sphere_closed_v7_stability_plot_forces_06022024_combined_v1_27022024} the components $F_r$ (dashed line) and $F_\theta$ (solid line) of the normalised hydrodynamic force exerted on the spindle, against angular position $\theta_\text{sp}$.
We indicate the slip angle $\theta_\text{slip}$ with green and the boundary of the stability region $\theta_\text{stable}$ in blue (also illustrated in Fig.~\ref{fig:fig_diagram_setup_v2_09112023}).

We see that our physical model is sufficient to produce stable spindle positioning, as observed in experiments~\cite{yi2011dynamic}.
In Fig.~\ref{fig:plot_41_45_force_fixed_sphere_closed_v7_stability_plot_forces_06022024_combined_v1_27022024}A, the fixed point $\theta_\text{sp}=0$ is stable; we choose the dimensionless spindle radius as $a=0.2$ (i.e.~ratio of spindle radius to oocyte radius) and active slip angle as $\theta_\text{slip}=1.31~\si{\radian}$~($=\ang{75}$), close to biological values. 
Near the fixed point, the hydrodynamic force pulls the spindle radially towards the cell cortex ($F_r\geq 0$), so the spindle remains near the cortex. 
Thus, only the $\theta$ dynamics matter, and the $\theta$ component of force pulls the spindle back to the fixed point ($F_\theta \leq 0$). 
Mathematically, we define the stability region for the fixed point $\theta_\text{sp}=0$ as the region $0\leq \theta_\text{sp} \leq \theta_\text{stable}$, where its size $\theta_\text{stable}$ is the largest $\theta_\text{sp}$ such that $F_r\geq 0$ and $F_\theta \leq 0$ for all $\theta_\text{sp}\leq \theta_\text{stable}$. 
This corresponds to the inset labelled ``stable" in Fig.~\ref{fig:plot_41_45_force_fixed_sphere_closed_v7_stability_plot_forces_06022024_combined_v1_27022024}A. 
In contrast with this, outside the stability region (i.e.~$\theta_\text{sp}>\theta_\text{stable}$), the spindle is pushed away from the cell cortex ($F_r<0$ as shown in ``unstable" inset in Fig.~\ref{fig:plot_41_45_force_fixed_sphere_closed_v7_stability_plot_forces_06022024_combined_v1_27022024}A). 

The fixed point is instead unstable for the parameter values used in Fig.~\ref{fig:plot_41_45_force_fixed_sphere_closed_v7_stability_plot_forces_06022024_combined_v1_27022024}B; the stability region has size $\theta_\text{stable}=0$ (blue).
Here, the dimensionless spindle radius is still $a=0.2$, but the active slip angle  $\theta_\text{slip}=2.36~\si{\radian}$~($=\ang{135}$) (green) is larger than for the stable case in Fig.~\ref{fig:plot_41_45_force_fixed_sphere_closed_v7_stability_plot_forces_06022024_combined_v1_27022024}A. 
Near the fixed point (i.e.~$\theta_\text{sp}$ small), even though the radial force pulls the spindle towards the cell cortex ($F_r\geq 0$), the $\theta$ component of force always drives the spindle away from the fixed point ($F_\theta\geq 0$ as in ``unstable" inset).

\subsection{Stability of spindle positioning depends critically on both spindle size and active slip angle at cell cortex}\label{sec:phase_diagram_spherical}

\begin{figure}[t]
	{\includegraphics[width=0.55\textwidth]{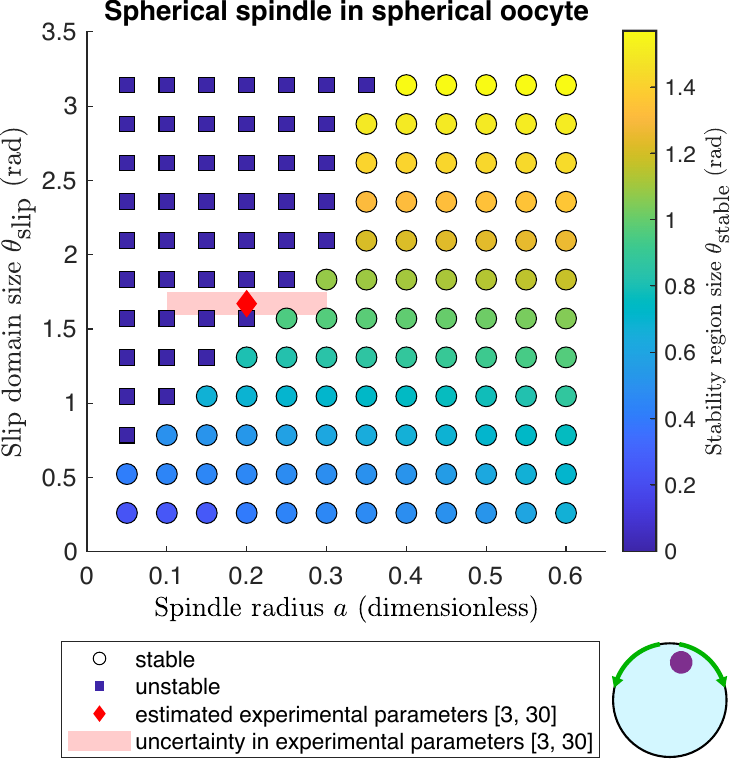}
		\caption{Phase diagram in $(a,\theta_\text{slip})$ plane showing stability of the fixed point $\theta_\text{sp}=0$ for the model spherical spindle of dimensionless radius $a$ inside an oocyte, with a prescribed active slip velocity on domain of angular size $\theta_\text{slip}$. 
			Circular markers indicate a stable fixed point, with colour showing the size of the stability region $\theta_\text{stable}$. 
			Square markers represent an unstable fixed point.
			Experimental parameters (red diamond marker) and uncertainty (light red shading) are estimated from Ref.~\cite{yi2011dynamic} and Ref.~\cite{niwayama2016bayesian}, as detailed in Sec.~\ref{sec:spindle_model}.
			 }
		\label{fig:plot_force_fixed_sphere_closed_v7_stability_phase_06022024_with_inset_v1_27022024}}
\end{figure}

Our model can produce  stable positioning of the spindle at the fixed point, for spindle radius and slip angle close to the biological values; however, for different parameter values, unstable spindle positioning is seen. 
The model is therefore able to capture the stability observed in experiments, but how robust is this stability?
We now explore the full parameter space numerically, to understand how the size of the stability region $\theta_\text{stable}$ depends on the spindle radius $a$ and the slip angle $\theta_\text{slip}$. 
We identify two key stability trends in terms of these parameters, which we will then explain physically with an analytical model.

Running many numerical simulations allows us to obtain the stability phase diagram in Fig.~\ref{fig:plot_force_fixed_sphere_closed_v7_stability_phase_06022024_with_inset_v1_27022024}; it shows the stability of the fixed point $\theta_\text{sp}=0$ as a function of $a$ and $\theta_\text{slip}$. 
Circular markers correspond to a stable fixed point, with colour indicating the size of the stability region $\theta_\text{stable}$, whereas square markers represent an unstable fixed point.  
The red diamond marker shows the estimated experimental parameters, with the uncertainty from our estimation of their values shaded in light red (see Sec.~\ref{sec:spindle_model} for details).

Remarkably, the range of experimental parameters lies at the boundary between stable and unstable spindle positioning in the phase diagram. 
While our model is idealised, designed to capture the essential  physics of the problem, it does predict that hydrodynamic effects alone are sufficient to keep the spindle at the fixed point; the cortex-driven flow may thus be fine-tuned to enable stable spindle positioning.  
Note that even if other biophysical mechanisms were shown to also contribute to the stable positioning of the spindle, hydrodynamics can provide a fail-safe for the oocyte (see further discussion in Sec.~\ref{sec:comparison_experiments}).

The phase diagram demonstrates two key features of the stability of hydrodynamic spindle positioning. 
First, for a given spindle radius, for sufficiently small slip angle, the spindle is stable at the fixed point.
Perhaps counter-intuitively, increasing the slip angle (i.e.~having a larger domain where fluid flow is locally directed away from the fixed point) increases the size of the stability region, until a further increase in the slip angle makes the fixed point unstable. 
We can understand this in terms of the forces shown in Fig.~\ref{fig:plot_41_45_force_fixed_sphere_closed_v7_stability_plot_forces_06022024_combined_v1_27022024}A~and~B. 
One consequence of a larger slip domain is a larger neighbourhood around the fixed point where the radial force $F_r$ is positive, which can increase the size of the stability region provided that $F_\theta$ is still negative. 
However, if the slip domain is sufficiently large, then the sign of $F_\theta$ near the fixed point can change from negative to positive, so stability is lost.

Thus, parameters close to the stability boundary in Fig.~\ref{fig:plot_force_fixed_sphere_closed_v7_stability_phase_06022024_with_inset_v1_27022024} that correspond to a stable fixed point also give rise to the largest possible stability region for a given spindle radius, indicating robustness of spindle positioning against perturbation; this applies to part of the experimental range of parameters (shaded in red). 

Secondly, at given slip angle, the fixed point is unstable for a small spindle and stable for a sufficiently large spindle. 
The intuitive physical picture of a local extensional flow systematically taking passively suspended particles away from the fixed point in Fig.~\ref{fig:diagram_motivating_spindle_stability}F is therefore only valid for sufficiently small particles; it is only because the spindle is large enough that it is hydrodynamically stable.

To explain the physical origin of these features, we next investigate the influence of confinement of the spindle inside the finite-sized cell on the stability of the fixed point, and discover that local hydrodynamic effects are sufficient to explain the stability.   
We conduct further simulations for a similar setup to that in Fig.~\ref{fig:fig_diagram_setup_v2_09112023}, but with a hole cut in the bottom of the cell,  within the no-slip region of the slip velocity.  
We find essentially the same stability results for this open cell as for the closed cell.
Therefore, keeping all other factors in the model unchanged, confinement is not necessary for stable spindle positioning. 
This motivates a second fundamental model, this time fully analytical and able to reveal the hydrodynamic mechanism, as we detail in the next section. 

\subsection{Hydrodynamic forces on spindle are captured by analytical model}\label{sec:force_simple}

In Sec.~\ref{sec:forces_spherical}, we showed that hydrodynamics can result in stable spindle positioning in a spherical model oocyte.
To understand the physical origin of this, we now introduce a minimal model that captures the essential physics of spindle positioning, reproducing analytically the stability results from our more detailed simulations. 
We demonstrate that stable spindle positioning can be understood in terms of a hydrodynamic suction (pulling) force, which originates from purely local hydrodynamic effects and systematically draws the spindle towards the fixed point.

\begin{figure}[t]
	{\includegraphics[width=0.75\textwidth]{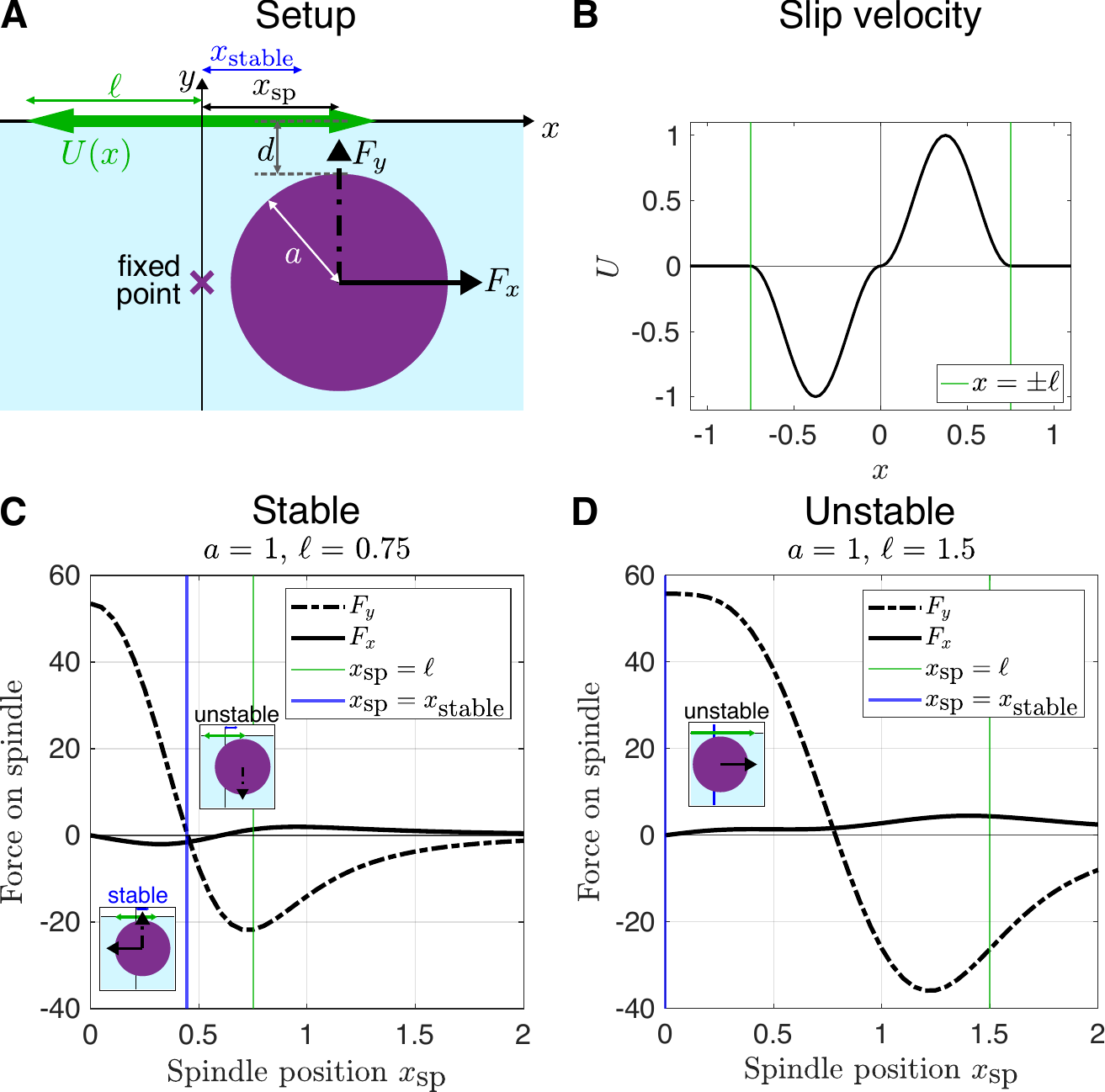} 
		\caption{
			(A and B)~Setup for simple analytical model, in two dimensions, to explain stability of spindle positioning at the fixed point.
			(A)~The spindle (purple) is modelled as a disc of radius $a$, near an active, planar wall at $y=0$, with clearance $d$. 
			A slip velocity boundary condition $u_x = U(x)$ with slip domain size $\ell$ (green in all panels) is prescribed on the $x$ axis, driving a flow in the fluid (light blue), which induces a hydrodynamic force on the spindle with components $F_x$ and $F_y$. 
			The fixed point (cross) is at $x=0$; the spindle position is $x_\text{sp}$.
			The stability region for the fixed point has size~$x_\text{stable}$ (blue in panels A, C and D). 
			(B)~Slip velocity $U(x)$ [Eq.~\eqref{eq:U_lub_slip_velocity}] vs $x$, for $\ell=0.75$. 
			(C~and~D)~Force on spindle [$F_y$ (dashed black) and $F_x$ (solid black)] against position~$x_\text{sp}$. 
			(C)~Stable spindle positioning at fixed point~$x_\text{sp}=0$ for $a=1$ and $\ell=0.75$.
			``Stable" inset: 
			in the stability region, with $x_\text{sp}$ positive~($0\leq x_\text{sp}\leq x_\text{stable})$, 
			the force satisfies $F_y \geq 0$ and $F_x \leq 0$; the fixed point is stable. 
			``Unstable" inset: $F_y<0$ outside the stability region (i.e.~$x_\text{sp} > x_\text{stable}$).
			(D)~Unstable spindle positioning at fixed point for  $a=1$ and  $\ell=1.5$. 
			``Unstable" inset:~$F_x > 0$.
		}
		\label{fig:fig_setup_fixed_spindle_near_cap_wall_lubrication_theory_v7_19042024_with_BC_v4_06052024_plot_28_64_spindle_fixed_circle_wall_lub_theory_plot_force_v7_23042024_with_inset_v1_23042024_combined_v3_13062024}}
\end{figure}

We illustrate the setup for our analytical model in Fig.~\ref{fig:fig_setup_fixed_spindle_near_cap_wall_lubrication_theory_v7_19042024_with_BC_v4_06052024_plot_28_64_spindle_fixed_circle_wall_lub_theory_plot_force_v7_23042024_with_inset_v1_23042024_combined_v3_13062024}A (see Sec.~\ref{sec:simple_physical_model_setup} for details). 
Working in two dimensions, we model the spindle as a rigid disc of radius $a$, near a planar, active wall along the $x$ axis that represents the cell cortex; the clearance $d$ is again assumed to be constant.
We denote the horizontal perturbation in spindle position from the fixed point by $x_\text{sp}$. 
As in Sec.~\ref{sec:forces_spherical}, the spindle is held stationary. 

On the planar wall, we prescribe a slip velocity boundary condition [Eq.~\eqref{eq:U_lub_slip_velocity}], which represents the forcing due to the actin-rich cortical cap of the oocyte. 
This is characterised by the maximum slip speed $U_\text{max}$ and the size of the active slip domain $\ell$ (analogous to the slip angle $\theta_\text{slip}$ in the spherical model in Sec.~\ref{sec:forces_spherical}).
We plot the slip velocity [Eq.~\eqref{eq:U_lub_slip_velocity}] in Fig.~\ref{fig:fig_setup_fixed_spindle_near_cap_wall_lubrication_theory_v7_19042024_with_BC_v4_06052024_plot_28_64_spindle_fixed_circle_wall_lub_theory_plot_force_v7_23042024_with_inset_v1_23042024_combined_v3_13062024}B, illustrating here with $\ell=0.75$. 
This boundary condition has symmetry about the line $x=0$; the origin therefore represents the centre of the cortical cap. 
The slip velocity is nonzero only inside the slip domain, i.e.~for $-\ell \leq x\leq \ell$, and drives a flow in the fluid below the cortex, modelling the motion of the cytoplasm (light blue in Fig.~\ref{fig:fig_setup_fixed_spindle_near_cap_wall_lubrication_theory_v7_19042024_with_BC_v4_06052024_plot_28_64_spindle_fixed_circle_wall_lub_theory_plot_force_v7_23042024_with_inset_v1_23042024_combined_v3_13062024}A).
In particular, $x_\text{sp}=0$ is a fixed point, since the spindle and the slip velocity share a line of symmetry in that case. 
We note that in all plots in Sec.~\ref{sec:force_simple} and Sec.~\ref{sec:phase_diagram_simple}, variables are nondimensionalised (details in Sec.~\ref{sec:simple_physical_model_setup}).

To analyse the stability of spindle positioning in this simple  model, we analytically solve for the fluid flow and pressure in the thin fluid film between the spindle and cell cortex, and hence the force on the spindle, using lubrication theory  (long-wavelength hydrodynamics). 
This means that our analytical model specifically captures all the local hydrodynamic effects that are involved in spindle positioning in our spherical model. 
Details of our assumptions, solution method and mathematical calculations are all provided in Sec.~\ref{sec:governing_simple} to Sec.~\ref{sec:horizontal_force}. 

To demonstrate stable and unstable spindle positioning in our simple analytical model, we now illustrate how the hydrodynamic force on the spindle varies with position of the spindle, for comparison with the numerical simulations of our spherical model in Sec.~\ref{sec:forces_spherical}. 
By symmetry of the slip velocity, we restrict our attention to $x_\text{sp} \geq 0$.
In Fig.~\ref{fig:fig_setup_fixed_spindle_near_cap_wall_lubrication_theory_v7_19042024_with_BC_v4_06052024_plot_28_64_spindle_fixed_circle_wall_lub_theory_plot_force_v7_23042024_with_inset_v1_23042024_combined_v3_13062024}C and D, we plot the vertical component $F_y$ [Eq.~\eqref{eq:Fy}] and horizontal component $F_x$ [Eq.~\eqref{eq:Fx}] of force on the spindle as a function of the spindle position $x_\text{sp}$. 

From the hydrodynamic forces, we can deduce the stability properties of the fixed point $x_\text{sp}=0$. 
In Fig.~\ref{fig:fig_setup_fixed_spindle_near_cap_wall_lubrication_theory_v7_19042024_with_BC_v4_06052024_plot_28_64_spindle_fixed_circle_wall_lub_theory_plot_force_v7_23042024_with_inset_v1_23042024_combined_v3_13062024}C, we see that the analytical model can produce stable spindle positioning at the fixed point $x_\text{sp}=0$; here, the spindle radius is $a=1$ and the size of the slip domain is $\ell=0.75$ (green).
Near the fixed point, when $x_\text{sp}$ is sufficiently small, the vertical hydrodynamic force on the spindle $F_y$ is positive, pulling the spindle towards the cell cortex.
Therefore, only the $x$ dynamics matter.
Furthermore, the horizontal force $F_x$ is negative, pulling the spindle back towards the fixed point (inset labelled ``stable"). 
The fixed point is thus stable. 
We define the stability region analogously to the spherical model case (Sec.~\ref{sec:forces_spherical}); its boundary $x_\text{sp}=x_\text{stable}$ is shown in blue in Fig.~\ref{fig:fig_setup_fixed_spindle_near_cap_wall_lubrication_theory_v7_19042024_with_BC_v4_06052024_plot_28_64_spindle_fixed_circle_wall_lub_theory_plot_force_v7_23042024_with_inset_v1_23042024_combined_v3_13062024}.
Mathematically, the stability region of the fixed point $x_\text{sp}=0$ is given by $-x_\text{stable} \leq x_\text{sp} \leq x_\text{stable}$, where $x_\text{stable}$ is the largest $x_\text{sp}$ such that $F_y \geq 0$ and $F_x\leq 0$ for all $x_\text{sp}$ with $0 \leq x_\text{sp} \leq x_\text{stable}$. 
Outside the stability region (i.e.~for $x_\text{sp} > x_\text{stable}$), the spindle is pushed away from the cortex, as the vertical force $F_y$ is negative (``unstable" inset in Fig.~\ref{fig:fig_setup_fixed_spindle_near_cap_wall_lubrication_theory_v7_19042024_with_BC_v4_06052024_plot_28_64_spindle_fixed_circle_wall_lub_theory_plot_force_v7_23042024_with_inset_v1_23042024_combined_v3_13062024}C).

For different parameter values, the model can instead predict  unstable spindle positioning at the fixed point $x_\text{sp}=0$, as shown in Fig.~\ref{fig:fig_setup_fixed_spindle_near_cap_wall_lubrication_theory_v7_19042024_with_BC_v4_06052024_plot_28_64_spindle_fixed_circle_wall_lub_theory_plot_force_v7_23042024_with_inset_v1_23042024_combined_v3_13062024}D. 
Here the spindle radius is the same as in the stable case ($a=1$), but the slip domain size is $\ell=1.5$, larger than in Fig.~\ref{fig:fig_setup_fixed_spindle_near_cap_wall_lubrication_theory_v7_19042024_with_BC_v4_06052024_plot_28_64_spindle_fixed_circle_wall_lub_theory_plot_force_v7_23042024_with_inset_v1_23042024_combined_v3_13062024}C.
For small $x_\text{sp}$, the spindle is pulled vertically towards the cortex ($F_y>0$).
Hence, just as for the stable fixed point in  Fig.~\ref{fig:fig_setup_fixed_spindle_near_cap_wall_lubrication_theory_v7_19042024_with_BC_v4_06052024_plot_28_64_spindle_fixed_circle_wall_lub_theory_plot_force_v7_23042024_with_inset_v1_23042024_combined_v3_13062024}C, only the horizontal dynamics matter. 
However, in Fig.~\ref{fig:fig_setup_fixed_spindle_near_cap_wall_lubrication_theory_v7_19042024_with_BC_v4_06052024_plot_28_64_spindle_fixed_circle_wall_lub_theory_plot_force_v7_23042024_with_inset_v1_23042024_combined_v3_13062024}D, we see that the horizontal force now pushes the spindle away from the fixed point ($F_x\geq 0$ in ``unstable" inset), which, consequently, is now unstable. 

\subsection{Analytical model reproduces stability trends of spherical model computations}\label{sec:phase_diagram_simple}

The force results shown in Fig.~\ref{fig:fig_setup_fixed_spindle_near_cap_wall_lubrication_theory_v7_19042024_with_BC_v4_06052024_plot_28_64_spindle_fixed_circle_wall_lub_theory_plot_force_v7_23042024_with_inset_v1_23042024_combined_v3_13062024} demonstrate that the simple analytical model produces the same physical features as the full spherical model solved numerically (Sec.~\ref{sec:forces_spherical}). 
To further compare the two, we explore the parameter space of the analytical model and find how the size of the stability region $x_\text{stable}$ depends on the spindle radius~$a$ and the active slip domain size~$\ell$.
 
\begin{figure}[t]
	{\includegraphics[width=0.55\textwidth]{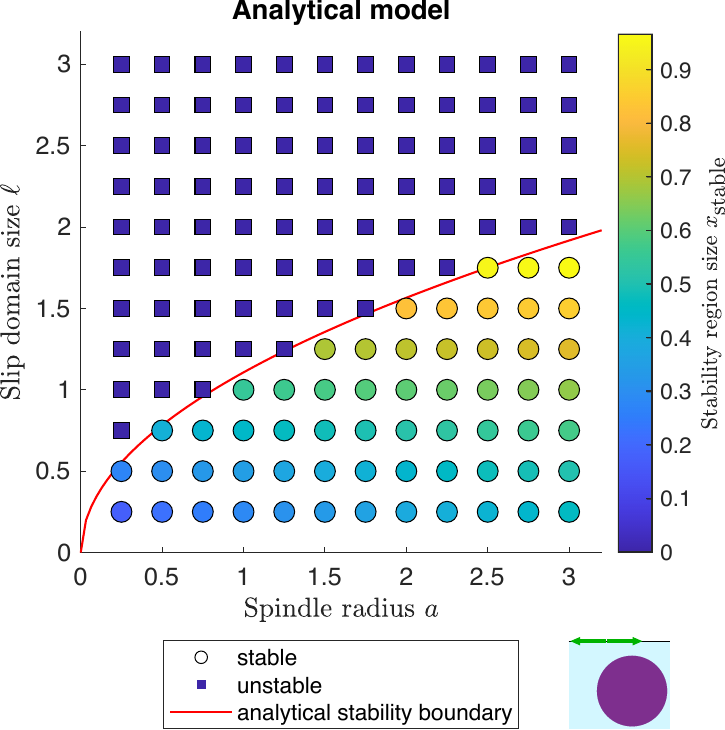}
		\caption{
			Phase diagram in the $(a,\ell)$ plane showing stability of the fixed point $x_\text{sp}=0$ in the analytical model, with the model spindle of radius $a$ near an active, planar wall with size of active slip domain $\ell$. 
			Circular markers represent a stable fixed point, with the stability region size $x_\text{stable}$ indicated by colour, while square markers correspond to an unstable fixed point.
			Red curve indicates analytical stability boundary [Eq.~\eqref{eq:stability_boundary_3}], from linear stability analysis.			
			}
		\label{fig:plot_1_stability_spindle_fixed_circle_wall_lub_theory_phase_w_bdry_v4_27102023}}
\end{figure}

In Fig.~\ref{fig:plot_1_stability_spindle_fixed_circle_wall_lub_theory_phase_w_bdry_v4_27102023}, we illustrate with a phase diagram in the~$(a,\ell)$ plane the stability results for our analytical model. As in Fig.~\ref{fig:plot_force_fixed_sphere_closed_v7_stability_phase_06022024_with_inset_v1_27022024}, 
circular markers represent a stable fixed point (here at $x_\text{sp}=0$), whereas square markers correspond to instability;  
colour indicates the stability region size.

Further, using linear stability analysis (detailed in Sec.~\ref{sec:linear_stability}), we can analytically derive the equation of the boundary between the stable and unstable regions of the phase diagram as
\begin{align}
	\ell = C d^{1/2} a^{1/2},\label{eq:stability_boundary_3_main}
\end{align}
where the dimensionless constant $C \approx 3.50$.
We plot this boundary in red, in perfect agreement with the results of the phase diagram. 

Importantly, with only local hydrodynamic effects, our simple analytical model reproduces the two key features in Sec.~\ref{sec:phase_diagram_spherical} of the phase diagram from the computational model (Fig.~\ref{fig:plot_force_fixed_sphere_closed_v7_stability_phase_06022024_with_inset_v1_27022024}) 
(note that the slip domain size is quantified by the parameter~$\ell$ for the analytical model but by the slip angle $\theta_\text{stable}$ in the computational model).  
First, at given spindle radius, increasing the size of the active slip domain $\ell$ makes the stability region larger, before making the fixed point unstable [for $\ell$ larger than that in Eq.~\eqref{eq:stability_boundary_3_main}]. 
Secondly, at a given slip domain size, for small spindle radius $a$, the fixed point is unstable, but for sufficiently large spindle radius, the fixed point is stable.

\subsection{Analytical model reveals hydrodynamic suction mechanism for stable spindle positioning}\label{sec:physical_mechanism}

What is the physical mechanism behind stable spindle positioning? Since the analytical model captures the physical features of simulations, we can use it to explain the hydrodynamic mechanism at the origin of the stable positioning. 
We thus examine in this section the flow, the pressure field and the contributions to the horizontal force $F_x$ exerted by the fluid on the spindle in the analytical model.
 
We first rationalise the hydrodynamic forces exerted on a spindle placed symmetrically, at the fixed point $x_\text{sp}=0$. 
We then consider how these are altered when the spindle is perturbed horizontally from the fixed point, for the two cases of stable and unstable spindle positioning.
We find that fluid flow can support stable spindle positioning by exerting a suction force on the spindle towards the fixed point. 

\subsubsection{Spindle at fixed point}

\begin{figure}[t] 
	\centering
	\includegraphics[width=0.75\textwidth]{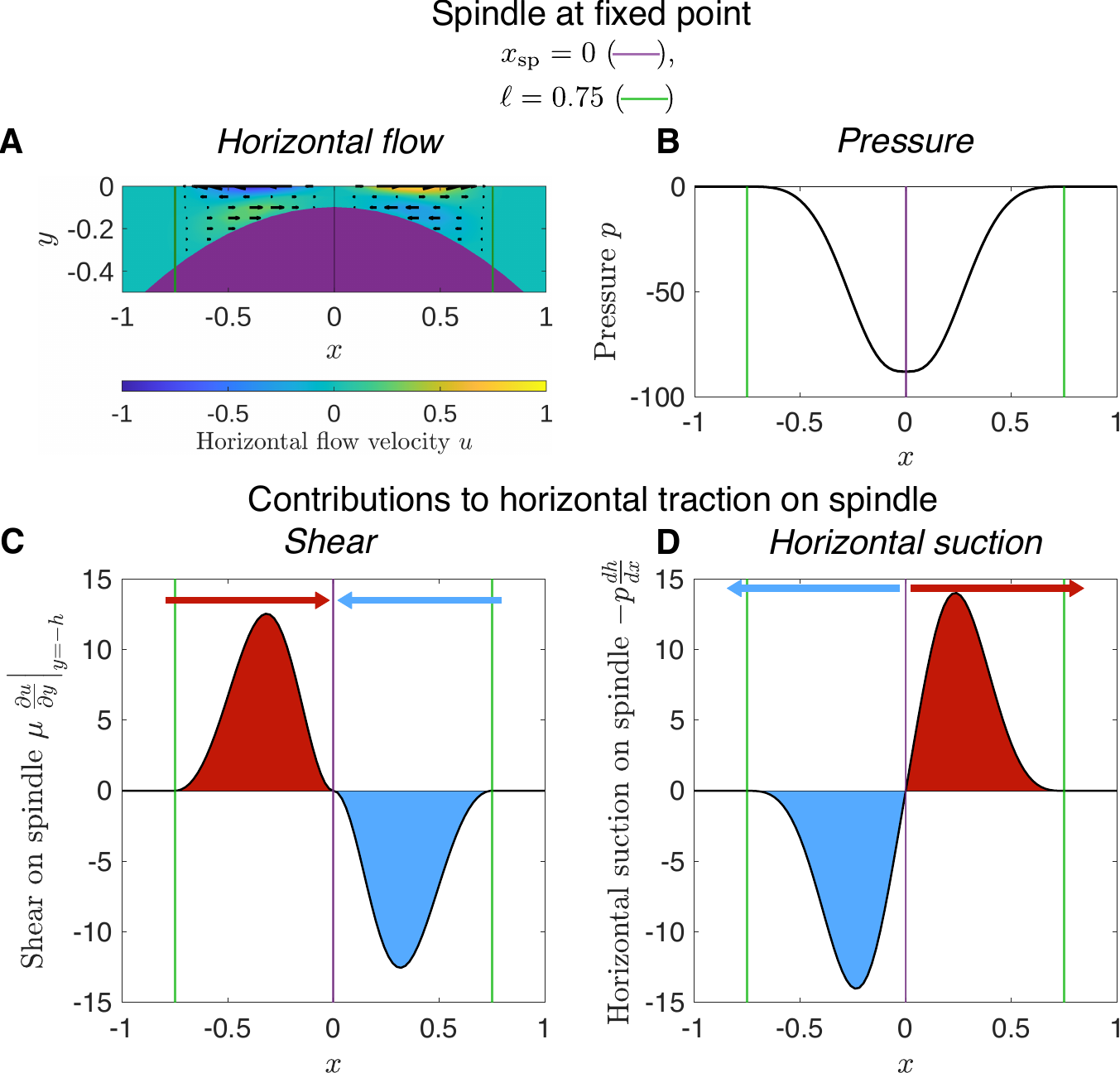}
	\caption{
		Illustration of physical mechanism for forces on spindle at the fixed point $x_\text{sp}=0$. Flow, pressure and contributions to horizontal traction (force per unit area) on symmetrically positioned spindle ($x_\text{sp}=0$, purple), with spindle radius $a=1$ and active slip domain size $\ell=0.75$ (green) (parameter values matching Fig.~\ref{fig:fig_setup_fixed_spindle_near_cap_wall_lubrication_theory_v7_19042024_with_BC_v4_06052024_plot_28_64_spindle_fixed_circle_wall_lub_theory_plot_force_v7_23042024_with_inset_v1_23042024_combined_v3_13062024}C). 
		(A)~Horizontal flow: arrows and colour indicate $u$.
		(B)~Pressure field as a function of $x$. 
		(C)~Shear on the spindle $\mu \left . \frac{\partial u}{\partial y} \right \vert_{y=-h}$ vs $x$. 
		(D)~Horizontal suction on the spindle $-p \frac{dh}{dx}$ vs $x$. 
		In panels C and D, the red shaded area and arrow indicate the  {contribution in the positive $x$ direction} to horizontal shear (C) or suction (D) force, while blue represents the negative contribution.
		}
	\label{fig:fig_plots_stable0_spindle_lub_understand_mechanism_v10_01022024_v2_12032024}
\end{figure}

We consider in Fig.~\ref{fig:fig_plots_stable0_spindle_lub_understand_mechanism_v10_01022024_v2_12032024} the case of a spindle positioned at the fixed point $x_\text{sp}=0$, i.e.~symmetrically with respect to the slip velocity. 
We use parameter values of spindle radius $a=1$ and slip domain size $\ell=0.75$ (green), the same as in Fig.~\ref{fig:fig_setup_fixed_spindle_near_cap_wall_lubrication_theory_v7_19042024_with_BC_v4_06052024_plot_28_64_spindle_fixed_circle_wall_lub_theory_plot_force_v7_23042024_with_inset_v1_23042024_combined_v3_13062024}C.

In Fig.~\ref{fig:fig_plots_stable0_spindle_lub_understand_mechanism_v10_01022024_v2_12032024}A, we show the horizontal flow with arrows, with the background colour indicating values of the horizontal flow velocity $u$.
The horizontal flow has a parabolic profile~[Eq.~\eqref{eq:u_lub}]. 
Near the cortex (i.e.~near the $x$ axis in the model), the direction of horizontal flow is inherited from the imposed slip velocity boundary condition; the flow is shear-dominated. 
However, closer to the spindle (purple), there is backflow (i.e.~flow in the opposite direction to the active slip velocity), conserving mass (here the mass flux $q$ is zero at $x=0$ by symmetry and hence zero everywhere).

We next focus on the pressure field, in Fig.~\ref{fig:fig_plots_stable0_spindle_lub_understand_mechanism_v10_01022024_v2_12032024}B. 
At the line of symmetry $x=0$, the pressure reaches its minimum and is negative. 
This shows that the flow, driven by the slip velocity, induces suction near $x=0$; the pressure is lower than in the absence of flow (i.e.~lower than the far-field value of zero). 
There are two immediate consequences of this. 
First, we see that the vertical hydrodynamic force exerted on the spindle [related to the pressure via Eq.~\eqref{eq:Fy}] is positive, pulling the spindle stably towards the cell cortex. 
Hence,  the sign of the horizontal force solely determines the linear stability of the spindle at the fixed point.
Secondly, there is a pressure gradient from infinity towards $x=0$, which drives the backflow near the spindle in Fig.~\ref{fig:fig_plots_stable0_spindle_lub_understand_mechanism_v10_01022024_v2_12032024}A required to conserve mass.

Equipped with this understanding of the flow and pressure fields, we can now explain the horizontal hydrodynamic forces exerted on the spindle.
In Fig.~\ref{fig:fig_plots_stable0_spindle_lub_understand_mechanism_v10_01022024_v2_12032024}C~and~D, we plot the two contributions to the horizontal traction (force per unit area) on the spindle [Eq.~\eqref{eq:Fx_suction_shear}]:  the shear on the spindle $\mu \left . \frac{\partial u}{\partial y} \right \vert_{y=-h}$ and the horizontal suction on the spindle $-p \frac{dh}{dx}$, respectively, against $x$, where $y=-h(x)$ describes the surface of the spindle near the cortex [Eq.~\eqref{eq:h_lub}]. 
Together, the integrals of these two contributions over $x$ (shaded area) give us the corresponding contributions to the horizontal force~$F_x$. 

The shear (Fig.~\ref{fig:fig_plots_stable0_spindle_lub_understand_mechanism_v10_01022024_v2_12032024}C) is seen to be rightwards on the left half of the spindle and leftwards on the right, as expected intuitively from inspecting the backflow near the spindle in Fig.~\ref{fig:fig_plots_stable0_spindle_lub_understand_mechanism_v10_01022024_v2_12032024}A. 
In contrast, the horizontal suction (Fig.~\ref{fig:fig_plots_stable0_spindle_lub_understand_mechanism_v10_01022024_v2_12032024}D) is leftwards on the left half of the spindle and rightwards on the right.
To understand this, we recall that the component of the local force due to the pressure acts perpendicularly to the spindle surface.
Since the pressure is negative, the force is a suction, thus pointing outwards from the spindle, towards and along the cortex.
Taking the horizontal component then gives us the directions of horizontal suction we see on each half of the spindle; geometrically, the deviation of the local slope of spindle surface from that of the cortex enables the spindle to experience a component of suction parallel to the cortex. 
By symmetry in $x=0$ (for $x_\text{sp}=0$), in Fig.~\ref{fig:fig_plots_stable0_spindle_lub_understand_mechanism_v10_01022024_v2_12032024}C, the rightward shear force (red shaded area and arrow) exactly balances the leftward (blue); similarly, the horizontal suction forces cancel out in Fig.~\ref{fig:fig_plots_stable0_spindle_lub_understand_mechanism_v10_01022024_v2_12032024}D. 
The total horizontal force on the spindle at the fixed point is thus zero, as expected. 

\subsubsection{Physical mechanism for stable and unstable spindle positioning}

\begin{figure}[t]  
	\centering
	\includegraphics[width=0.88\textwidth]{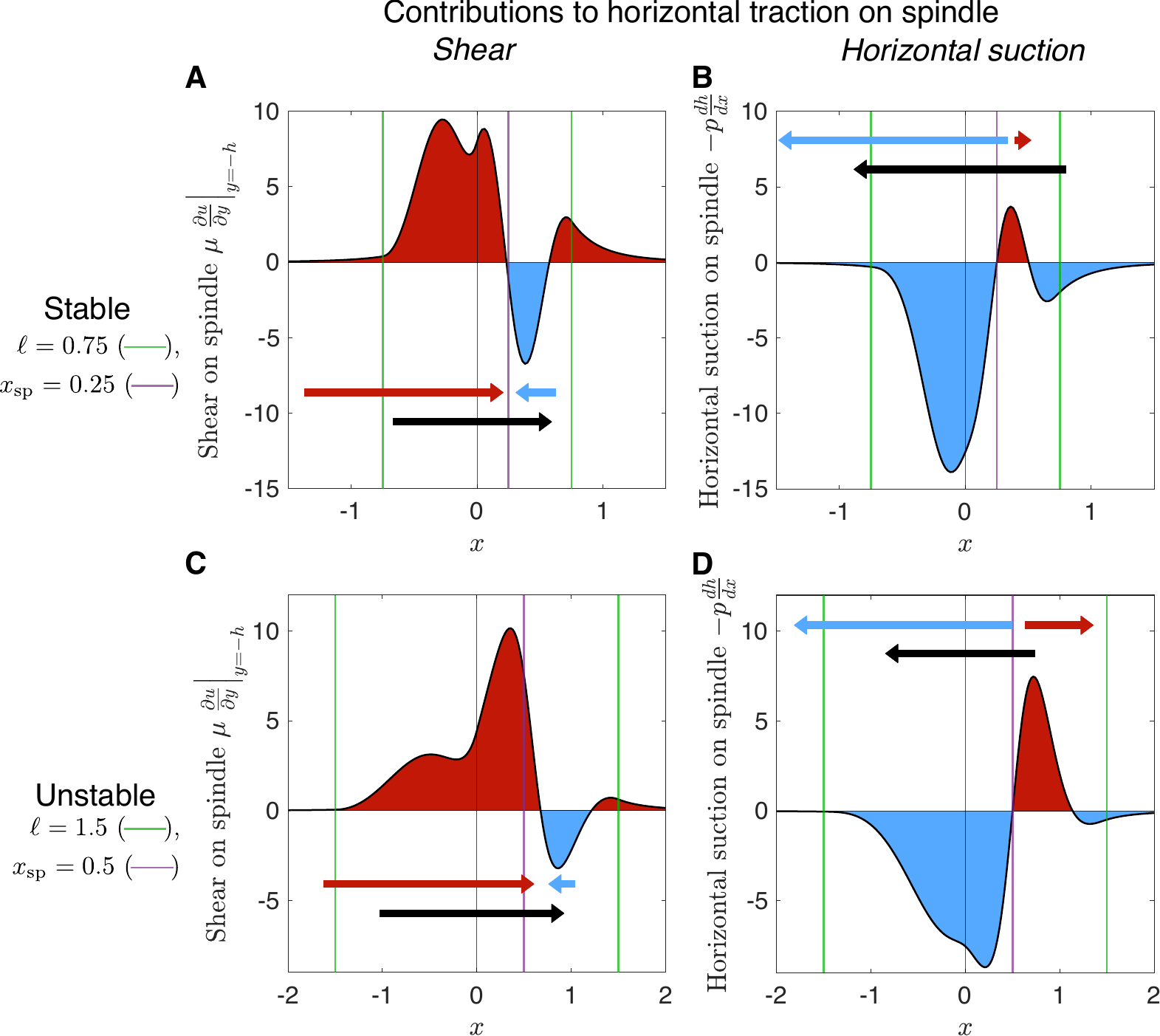}
	\caption{Physical mechanism behind stable and unstable spindle positioning. 
		Contributions to horizontal traction on the spindle positioned just to the right of the fixed point at $x=0$, for stable (top row; $a=1$, $\ell=0.75$, as in Fig.~\ref{fig:fig_setup_fixed_spindle_near_cap_wall_lubrication_theory_v7_19042024_with_BC_v4_06052024_plot_28_64_spindle_fixed_circle_wall_lub_theory_plot_force_v7_23042024_with_inset_v1_23042024_combined_v3_13062024}C) and  unstable cases (bottom row;  $a=1$,  $\ell=1.5$, as in Fig.~\ref{fig:fig_setup_fixed_spindle_near_cap_wall_lubrication_theory_v7_19042024_with_BC_v4_06052024_plot_28_64_spindle_fixed_circle_wall_lub_theory_plot_force_v7_23042024_with_inset_v1_23042024_combined_v3_13062024}D).
		Green indicates boundaries of the slip domain $x=\pm\ell$. 
		Purple shows  spindle position $x=x_\text{sp}$, with $x_\text{sp}=0.25$ in the top row and $x_\text{sp}=0.5$ in the bottom row. 
		(A~and~C)~Shear on the spindle $\mu \left . \frac{\partial u}{\partial y} \right \vert_{y=-h}$ as a function of~$x$. 
		(B~and~D)~Horizontal suction on the spindle~$-p \frac{dh}{dx}$ vs~$x$. 
		As in Fig.~\ref{fig:fig_plots_stable0_spindle_lub_understand_mechanism_v10_01022024_v2_12032024}, red shaded areas and arrows represent positive contributions to horizontal force~{$F_x$}, whereas blue indicates negative contributions; each arrow has length proportional to the shaded area.
		The net contribution to horizontal force (i.e.~red plus blue) is shown with a black arrow.
	}
	\label{fig:fig_plots_stable_unstable_spindle_lub_mechanism_v10_01022024_v2_12032024}
\end{figure}

Having rationalised the forces on a spindle positioned exactly at the fixed point ($x_\text{sp}=0$), we next turn our attention to the effect of a small perturbation in the spindle position, for example with $x_\text{sp} >0$. 
This analysis of the horizontal spindle dynamics near the fixed point will highlight the key physical result in our work. 
 
In Fig.~\ref{fig:fig_plots_stable_unstable_spindle_lub_mechanism_v10_01022024_v2_12032024}, we illustrate how the forces experienced by the spindle are modified by the small perturbation to the right, for parameter values corresponding to a stable fixed point in the top row (spindle radius $a=1$, slip domain size $\ell=0.75$, as in Fig.~\ref{fig:fig_setup_fixed_spindle_near_cap_wall_lubrication_theory_v7_19042024_with_BC_v4_06052024_plot_28_64_spindle_fixed_circle_wall_lub_theory_plot_force_v7_23042024_with_inset_v1_23042024_combined_v3_13062024}C) and to an unstable fixed point in the bottom row ($a=1$, $\ell=1.5$, as in Fig.~\ref{fig:fig_setup_fixed_spindle_near_cap_wall_lubrication_theory_v7_19042024_with_BC_v4_06052024_plot_28_64_spindle_fixed_circle_wall_lub_theory_plot_force_v7_23042024_with_inset_v1_23042024_combined_v3_13062024}D).
The active slip domain size $\ell$ is indicated in green and the spindle position ($x=x_\text{sp}$) in purple.
The two contributions to the horizontal traction on the spindle, shear $\mu \left . \frac{\partial u}{\partial y} \right \vert_{y=-h}$ and horizontal suction $-p \frac{dh}{dx}$, are shown in the left~(Fig.~\ref{fig:fig_plots_stable_unstable_spindle_lub_mechanism_v10_01022024_v2_12032024}A~and~C) and right~(Fig.~\ref{fig:fig_plots_stable_unstable_spindle_lub_mechanism_v10_01022024_v2_12032024}B~and~D) columns, respectively.
As in Fig.~\ref{fig:fig_plots_stable0_spindle_lub_understand_mechanism_v10_01022024_v2_12032024}, the red and blue shaded areas represent the positive and negative contributions 
to horizontal force~$F_x$, respectively; arrow length is proportional to the total shaded area of the same colour.
Further, the net contribution to horizontal force from each physical effect is shown with a black arrow (i.e.~the sum of red and blue).

To understand the influence of the perturbation of spindle position and hence the stability of the fixed point at $x=0$, we compare the horizontal forces with those for the spindle positioned at the fixed point, examining specifically Fig.~\ref{fig:fig_plots_stable_unstable_spindle_lub_mechanism_v10_01022024_v2_12032024}A~and~C vs Fig.~\ref{fig:fig_plots_stable0_spindle_lub_understand_mechanism_v10_01022024_v2_12032024}C~(shear), and Fig.~\ref{fig:fig_plots_stable_unstable_spindle_lub_mechanism_v10_01022024_v2_12032024}B~and~D vs Fig.~\ref{fig:fig_plots_stable0_spindle_lub_understand_mechanism_v10_01022024_v2_12032024}D~(horizontal suction). 
First, for the stable fixed point, in Fig.~\ref{fig:fig_plots_stable_unstable_spindle_lub_mechanism_v10_01022024_v2_12032024}A, we find that the rightward, destabilising shear force (red) now outweighs the leftward, stabilising contribution (blue), so the net shear force is rightwards and destabilising (black arrow). 
In contrast with this, we see in Fig.~\ref{fig:fig_plots_stable_unstable_spindle_lub_mechanism_v10_01022024_v2_12032024}B that the leftward, stabilising horizontal suction (blue) dominates over the rightward suction (red); hence, the net horizontal suction force is leftwards and stabilising (black arrow). 
Importantly, the stabilising net horizontal suction force is greater than the destabilising net shear force, so the total horizontal force on the spindle is stabilising, i.e.~$F_x<0$ as we see in Fig.~\ref{fig:fig_setup_fixed_spindle_near_cap_wall_lubrication_theory_v7_19042024_with_BC_v4_06052024_plot_28_64_spindle_fixed_circle_wall_lub_theory_plot_force_v7_23042024_with_inset_v1_23042024_combined_v3_13062024}C at $x_\text{sp}=0.25$. 
In essence, the low pressure leads directly to a hydrodynamic suction pulling spindle back to the fixed point; the finite size of the spindle enables it to detect the stabilising pressure difference across its length. 
This mechanism differs fundamentally from hydrodynamic trapping of particles near stagnation points in other physical regimes, where, for example (and unlike in the oocyte), inertia can play an important role~\cite{raju1997dynamics,ramarao1989aerosol,zhong2007dynamical}.

In the case of an unstable fixed point, Fig.~\ref{fig:fig_plots_stable_unstable_spindle_lub_mechanism_v10_01022024_v2_12032024}C and D illustrate similar behaviour to Fig.~\ref{fig:fig_plots_stable_unstable_spindle_lub_mechanism_v10_01022024_v2_12032024}A and B, respectively. 
However, the crucial difference in this case is that the destabilising net shear force outweighs the stabilising net horizontal suction force.
Hence, the spindle is now destabilised by the total horizontal force, i.e.~$F_x>0$ (as in Fig.~\ref{fig:fig_setup_fixed_spindle_near_cap_wall_lubrication_theory_v7_19042024_with_BC_v4_06052024_plot_28_64_spindle_fixed_circle_wall_lub_theory_plot_force_v7_23042024_with_inset_v1_23042024_combined_v3_13062024}D at $x_\text{sp}=0.5$).
Here, intuitively, a spindle that is too small compared with the active slip domain acts like a particle; its smaller relative size reduces its ability to experience pressure gradients in the flow, so the horizontal suction on the two halves of the spindle largely cancel out and the spindle is pushed away from the fixed point by the flow. 

The key to understanding the hydrodynamic trapping mechanism for the spindle is thus the difference between the two rows in Fig.~\ref{fig:fig_plots_stable_unstable_spindle_lub_mechanism_v10_01022024_v2_12032024}.

\section{Discussion}\label{sec:discussion}

\subsection{Summary}

We showed in this article that the stability of spindle positioning just beneath the cortical cap, while an oocyte awaits fertilisation, can be explained by purely local hydrodynamic effects.
We saw in Sec.~\ref{sec:introduction} that although the spindle in a meiosis II oocyte remains stably positioned at a fixed point for many hours, the fixed point intuitively appears  to be unstable in the experimentally measured cytoplasmic flow field.
To explain this observed stability, in Sec.~\ref{sec:flow_simulation}, we introduced a physical model for the flow inside the cell, driven by an active slip velocity and based on experimental data~\cite{niwayama2016bayesian,yi2011dynamic}.
We then presented our model for the spindle in Sec.~\ref{sec:forces_spherical}, identifying the key parameters of our model as the radius of the spherical spindle and the size of the slip domain at the cortex. 
To understand quantitatively the stability of the fixed point, we computed numerically the force on the spindle as a function of its position. 
We found that our model is able to reproduce the stability of the fixed point observed experimentally and that, exploring the parameter space in Sec.~\ref{sec:phase_diagram_spherical}, the stability of the fixed point depends critically on the spindle radius and the slip domain size: for a given spindle size, the fixed point is stable for small slip angle and unstable for large slip angle; for a given slip angle, the fixed point is stable for a sufficiently large spindle.
Next, to reveal the mechanism underlying this, we presented in Sec.~\ref{sec:force_simple} a simple analytical model that captures the essential ingredients of the spherical model in Sec.~\ref{sec:forces_spherical}. 
We focused on the fluid flow between the spindle and the cell cortex, i.e.~local hydrodynamics.
We calculated analytically the stability properties of the fixed point as a function of spindle radius and size of the slip domain in Sec.~\ref{sec:phase_diagram_spherical}, reproducing the trends we found for the spherical model.
Using this analysis, in Sec.~\ref{sec:physical_mechanism}, we explained the stability of the fixed point in terms of two competing effects induced by the flow: the shear (always leading to a destabilising force exerted on the spindle) and the horizontal suction on the spindle (always stabilising). 
When the spindle is sufficiently large relative to the slip domain, the stabilising suction force can overcome the destabilising shear. 
Cytoplasmic fluid flow can thus be sufficient to keep the spindle stably positioned at the fixed point, just beneath the cortical cap of an oocyte.

\subsection{Comparison with experiments}\label{sec:comparison_experiments}

In our study, we examined the stability of spindle positioning within the framework of a hydrodynamic model. 
We found that horizontal suction due to the fluid flow inside the cell could explain the stability of the fixed point, making the system robust against perturbations. 
However, although hydrodynamics can be sufficient according to our theoretical modelling, other physical, chemical and biological ingredients could also play a part in stable spindle positioning.
We showed theoretically the condition under which the cytoplasmic flow by itself could hold the spindle in place, revealing a mechanism for hydrodynamic stabilisation. 
Thus, in an oocyte, cytoplasmic flow could be either beneficial for or detrimental to stable spindle positioning, depending on parameter values. 
If it is indeed beneficial, then cytoplasmic flow could either provide the dominant mechanism for stable spindle positioning, via the physics in this article, or act as a fail-safe if another mechanism is primarily responsible. 
If, instead, cytoplasmic flow is detrimental, our model can be used to provide information on the magnitude of the mechanism that stabilises the spindle against this hydrodynamic effect.

First, let us consider the role of actin~\cite{yi2012actin}.
We have so far discussed its flow-mediated effect on stable spindle positioning, in the context of Ref.~\cite{yi2011dynamic}, which showed experimentally that actin flow drives the cytoplasmic streaming that we analysed theoretically in our work.
Studies prior to this investigation of flow in Ref.~\cite{yi2011dynamic} had also demonstrated that actin is important for spindle positioning in meiosis II oocytes~\cite{azoury2008spindle,longo1985development,maro1986mechanism,halet2007rac}.
In particular, attachment of the spindle to the cortex, mediated by actin microfilaments, had been suggested~\cite{halet2007rac}. 
Subsequently, the experiments of Ref.~\cite{yi2011dynamic} revealed a mechanism by which actin could dynamically keep the spindle near the cortex via cytoplasmic streaming, as we outlined in Sec.~\ref{sec:introduction}. 

We wish to highlight evidence from two types of experiment relevant to our theoretical model.
First, in past experimental work, the spindle was observed to migrate from the interior of the oocyte towards the fixed point, in a way that was consistent with the fluid flow field~\cite{yi2011dynamic}.
In this, the authors of Ref.~\cite{yi2011dynamic} applied a treatment to oocytes, so that the cytoplasmic streaming driven by actin flow was deactivated temporarily and the spindle moved slightly away from the cortex, towards the interior of the oocyte.
They then stopped the treatment. 
Cytoplasmic streaming resumed and the spindle moved back towards the cortex at a similar speed to nearby cytoplasmic particles.
During this period of time, the spindle therefore moved towards the fixed point in a manner consistent with the fluid flow driven from the cell cortex and the suction force on the spindle predicted by our model.

Another notable experiment involves disassembly of the spindle (by treatment with an agent called nocodazole)~\cite{yi2011dynamic,maro1986mechanism}.
In terms of our fluid dynamical theory, this effectively replaces the  spindle with a smaller object inside the oocyte, which we expect could in turn impact its stability at the fixed point.
The authors of Ref.~\cite{yi2011dynamic} found that when the spindle was disassembled, the now-naked chromosomes (previously attached to the spindle) moved closer to the cortex, with the same speed as the flow of cytoplasmic particles nearby.
The chromosomes then remained there; we note that the experimental video shows their behaviour for a time period of around half an hour. 
This is consistent with the idea that flow typically contributes to the positioning of the spindle inside the cell.
Quantitatively, we may estimate the dimensionless radius of the cluster of chromosomes relative to the oocyte radius as approximately~$0.1$ and apply our theory to the cluster, instead of an intact spindle. 
Assuming the same active slip angle for the mouse oocyte as in our study, this does correspond to an unstable fixed point, according to the phase diagram in Fig.~\ref{fig:plot_force_fixed_sphere_closed_v7_stability_phase_06022024_with_inset_v1_27022024}. 
However, we estimate the timescale over which the chromosomes move away from the fixed point to be longer than the video duration: even if the cluster moved as fast as the maximum active slip velocity $U_\text{max} \approx 2~\si{\nano\metre}~\si{\second}^{-1}$, it would take $\approx 50$~minutes to travel $6~\si{\micro\metre}$ (i.e.~approximately a spindle radius). 
Therefore, despite the predicted unstable positioning, we may not expect it to be apparent in the aforementioned video.

In Ref.~\cite{maro1986mechanism}, after the dissolution of the spindle (again with nocodazole), over the course of several hours, the chromosomes dispersed; small clusters of chromosomes were redistributed around the cortex of the oocyte.
This is consistent with the prediction from our hydrodynamic model that for a given active slip domain size, the fixed point is unstable for smaller bodies (here  applying our results to individual clusters).
For example, an observed cluster~\cite{maro1986mechanism} with dimensionless radius of approximately $0.04$ relative to the oocyte radius (several times smaller than an intact spindle) lies in the unstable region of the phase diagram, if we again assume the same active slip angle.
Hence, cytoplasmic streaming generated by actin flow could be fine-tuned to enable stable positioning of the spindle purely by flow. 

Other past work has focused on the role of myosin (a molecular motor~\cite{mogessie2018assembly}) in spindle positioning~\cite{yi2011dynamic,bourdais2023mrck,mcginnis2015mapk3,schuh2008new}.
Experiments have demonstrated that inhibiting myosin II does not affect the actin flow or cytoplasmic streaming illustrated in Fig.~\ref{fig:diagram_motivating_spindle_stability}~\cite{yi2011dynamic}, and does not cause the spindle to drift away from the cortex~\cite{yi2011dynamic,bourdais2023mrck}.
Other groups have discussed the effect of different inhibitors and timescales for treatment of the oocytes~\cite{mcginnis2015mapk3,schuh2008new}.

We also note that microtubules are critical for spindle positioning in many cells~\cite{siller2009spindle,yi2011dynamic,wu2024laser}, including for meiosis I mouse oocytes~\cite{xie2018poly}. 
However, experimental work where microtubules were disrupted suggests that they are not responsible for asymmetric spindle positioning in the meiosis II mouse oocyte~\cite{siller2009spindle,yi2011dynamic}. 

\subsection{Limitations of theoretical modelling}

The purpose of the modelling presented in this paper is to demonstrate a physical mechanism that can keep the spindle at the fixed point and to understand its physical origin. 
The computational model  strikes a balance between realistic on one side (containing enough  ingredients to reproduce an experimental situation) and simplified on the other (being sufficiently idealised that a full theoretical analysis can be carried out). 

In a real oocyte, the boundary between stability and instability of the fixed point in parameter space may  quantitatively deviate from that predicted by our mathematical model in Sec.~\ref{sec:phase_diagram_spherical}.
Various properties of the spindle and its environment could thus be revisited in future work. 
First, in an oocyte, the shape of the spindle is approximately ellipsoidal instead of spherical. 
However, our simulations with an ellipsoidal model spindle produced similar behaviour to that for a spherical spindle shown in Sec.~\ref{sec:phase_diagram_spherical}. 
Furthermore, although we describe the spindle in our analytical model setup as circular (as a natural simplification of the spherical model), the results hold for more general spindle shapes (e.g.~elliptical) too. 
This is because we approximate the circle with a parabolic height profile locally, as is done generically in fluid dynamics problems in thin gaps; this quadratic profile is the leading-order approximation to a general, smooth spindle shape that is symmetrical about the normal to the model cortex, valid in the region where the spindle is closest to the cortex and which provides the dominant contribution to force.
The material properties of the spindle could also be modelled more realistically, for instance, by treating the  spindle as a flexible~\cite{oriola2018physics} and permeable~\cite{yi2011dynamic} structure. 
Further, our model does not include any active noise; the spindle is an active system, so would be expected to fluctuate~\cite{oriola2018physics}. 
In our work, as a first approximation, we treat the cytoplasm as a Newtonian fluid, with its viscosity modelling the resistance of the cytoplasm, but a non-Newtonian rheology~\cite{fle2023imaging} as well as the heterogeneous nature of the cytoplasm~\cite{shamipour2021cytoplasm} could also contribute to spindle positioning near the cell cortex~\cite{najafi2023size}. 

Despite these simplifications, our fundamental modelling was able to reproduce the experimental features of living oocytes and explain hydrodynamically why the spindle is expected to be stable. 
We hope that it will motivate further, more detailed biophysical modelling. 

\subsection{Outlook}

Further experiments would allow us to test our theory and assess to what extent flow is responsible for stable spindle positioning.
How important is flow in comparison with other potential mechanisms for spindle positioning? 
Complementary to the experimental study of Ref.~\cite{yi2011dynamic}, our work quantitatively predicts the stability of the fixed point and size of the stability region as a function of the two key parameters of the system, the ratio of spindle radius to oocyte radius and the active slip angle. 
We envisage that experimentalists could develop methods to explore this parameter space for living oocytes; 
perhaps the active slip angle could be modified via changes to the size of the actin cap. 
Experiments examining the effect of genetic and chemical perturbations on flow and actin dynamics~\cite{almonacid2014actin,nikalayevich2024aberrant,dehapiot2013polarized,bourdais2021cofilin,balzer2019single,lee2013thioredoxin,namgoong2016roles,wang2019rab23} could contribute to answering this question. 
Furthermore, imaging of actin and measurement of cytoplasmic flow specifically in the region between the spindle and cortex could also yield evidence on the mechanisms at play.

In oocytes of mammalian species beyond the mouse model, while the geometrical parameters of our model may be measured from existing images of the meiotic spindle~\cite{liu2000reliable}, new studies investigating cytoplasmic streaming in other species are needed to determine whether it is a conserved phenomenon. 
We note that a well-defined actin cap, localised to the cortex adjacent to the spindle, has been shown to be present in human meiosis II oocytes~\cite{coticchio2014contributions}. 
Then, with 
data on the cytoplasmic flow field and spindle size for more species of mammalian oocytes, we could place more experimental data points on our phase diagram in Sec.~\ref{sec:phase_diagram_spherical}; their position relative to the stability boundary would provide evidence on the role of hydrodynamics in other mammals. 

Future modelling could incorporate the discovered feedback loop between the location of the actin cap and the spindle position (in mouse oocytes)~\cite{yi2013symmetry}. 
In terms of stability of spindle positioning perpendicular to the cortex, the proximity of the spindle to the cortex has been shown to induce an actin cap at that location, via signals from the chromosomes~\cite{longo1985development,deng2007ran,maro1986mechanism,sun2011arp2}. 
The resulting actin flow and hence cytoplasmic streaming keep the spindle near the cortex~\cite{yi2013symmetry,yi2011dynamic}.
Turning our attention to perturbations parallel to the cortex, if the spindle moves along the cortex, then the actin cap could conceivably relocate to the spindle's new position. 
We would thus envision that the stagnation point in the cytoplasmic flow and hence the fixed point of spindle position would also move there. 
This could impact the stability of spindle positioning.
Depending on the timescale of actin cap formation in comparison with spindle motion, we speculate that a physical model including the feedback loop could exhibit rich behaviour, where the spindle could potentially return robustly to a stable region instead of merely a single stable fixed point.

The role of flow in stable spindle positioning in different stages of cell division~\cite{totsuka2024ca2+,wang2020symmetry,dehapiot2021rhoa} or in other species~\cite{mcnally2013mechanisms,li2013road,coticchio2014contributions} could also be explored. 
Cytoplasmic flow in oocytes can impact organelle distribution~\cite{nikalayevich2024aberrant,lu2023go}, and spindle positioning and orientation~\cite{totsuka2024ca2+,delgado2020mechanical}, and hence development~\cite{almonacid2018control}. 
Future work could therefore investigate whether flow can be used as an indicator of oocyte quality and female fertility~\cite{wasielak2022chromosome,babayev2022age}. 

\section{Methods}

\subsection{Cytoplasmic flow: model and numerical simulations}\label{sec:flow_model}

Our model setup for the cytoplasmic streaming is shown in Fig.~\ref{fig:plot_Niwayama_2016_least_squares_nlinfit_plot_v4_27022024_with_inset_v1_27022024}A (see also Sec.~\ref{sec:flow_simulation}).
Here, we use spherical polar coordinates $(r,\theta,\phi)$, where $r$ is the radial distance from the centre of the spherical model oocyte, $\theta$ is the polar angle and $\phi$ is the azimuthal angle. 
Mathematically, the intracellular fluid flow in our model is governed by the incompressible Stokes equations,
\begin{align}
	\mu \nabla^2 \mathbf{u} &= \nabla p,\\
	\nabla \cdot \mathbf{u} &= 0,
\end{align}
where $\mathbf{u}$ is the fluid velocity field with components $(u_r,u_\theta,u_\phi)$ in spherical polar coordinates, $p$ is the pressure field and $\mu$ is the dynamic viscosity of the fluid~\cite{kimbook}.  
The cell cortex is given by $r=r_\text{o}$, where $r_\text{o}$ is the radius of the oocyte. 
Fluid occupies the whole region inside the sphere (light blue in inset of Fig.~\ref{fig:plot_Niwayama_2016_least_squares_nlinfit_plot_v4_27022024_with_inset_v1_27022024}A).
Based on the experimental data in Fig.~\ref{fig:plot_Niwayama_2016_least_squares_nlinfit_plot_v4_27022024_with_inset_v1_27022024}A, we thus prescribe in our model the active slip velocity $u_\theta=U(\theta)$, with $u_r=u_\phi=0$, on the cortex $r=r_\text{o}$ as 
\begin{align}
	U(\theta) = 
	\begin{cases}
		U_\text{max} \sin^2\left (\frac{\pi \theta}{\theta_\text{slip}}\right ), & 0 \leq \theta \leq \theta_\text{slip},\\
		0, & \theta_\text{slip} < \theta \leq \pi,
	\end{cases}
	\label{eq:U_theta_slip_velocity}
\end{align}
where $U_\text{max}$ is the characteristic velocity scale and $\theta_\text{slip}$ is the slip angle.
The slip velocity is positive for $\theta\leq\theta_\text{slip}$ and zero elsewhere. 

The numerical simulation of flow (Fig.~\ref{fig:plot_Niwayama_2016_least_squares_nlinfit_plot_v4_27022024_with_inset_v1_27022024}B) was conducted in COMSOL Multiphysics® 5.6~\cite{COMSOL} using the physics-controlled, fine mesh.

\subsection{Spindle dynamics: spherical model and numerical simulations}\label{sec:spindle_model}

We illustrate our model for the spindle dynamics in Fig.~\ref{fig:fig_diagram_setup_v2_09112023} (see also Sec.~\ref{sec:forces_spherical}).
The radial position of the centre of the spherical model spindle is given by $r=r_\text{sp}\equiv r_\text{o}-d-a$, where we recall that $r_\text{o}$ is the oocyte radius, $d$ is the clearance and $a$ is the spindle radius. 
In each numerical simulation (COMSOL Multiphysics® 5.6~\cite{COMSOL} with physics-controlled, normal mesh),
the spindle is held stationary at a fixed position given in spherical polar coordinates by $(r=r_\text{sp}, \theta=\theta_\text{sp}, \phi=0)$, in the Stokes flow of fluid occupying the eccentric spherical annular region between the spindle and cell cortex (shaded light blue in Fig.~\ref{fig:fig_diagram_setup_v2_09112023}), driven by the prescribed active slip velocity in Eq.~\eqref{eq:U_theta_slip_velocity}.  
We assume the dimensionless clearance, defined as $\hat{d} \equiv d/r_\text{o}$, to be a constant, $\hat{d}=0.1$, measured from Fig.~\ref{fig:diagram_motivating_spindle_stability}B (reproduced from Ref.~\cite{yi2011dynamic}); in our model, a non-hydrodynamic repulsive force from the cortex is thus assumed to balance the radial hydrodynamic force attracting the spindle towards the cortex in a neighbourhood of the fixed point. 
On the spindle, we prescribe the no-slip boundary condition $\mathbf{u}=\mathbf{0}$. 
It may be verified that in Stokes flow, the $\theta$ component of velocity of a free model spindle (with constant clearance $d$) has the same sign as the hydrodynamic force component $F_\theta$, if the free spindle remains oriented parallel to the cortex; hence, the sign of the force indicates the direction in which the spindle would move if free to do so.

For the experimental parameters indicated in Fig.~\ref{fig:plot_force_fixed_sphere_closed_v7_stability_phase_06022024_with_inset_v1_27022024}, we obtain a biological value of the active slip angle $\theta_\text{slip}$ from the nonlinear least squares method on experimental data~\cite{niwayama2016bayesian}, as detailed in Sec.~\ref{sec:flow_simulation}; the uncertainty shown is the 95\% confidence interval for the fit of the model to the data. 
We measure the size of the meiosis II spindle from Fig.~\ref{fig:diagram_motivating_spindle_stability}B (reproduced from Ref.~\cite{yi2011dynamic}) and treat this spindle as a sphere of dimensionless radius~$a=0.2 \pm 0.1$~(i.e.~measured relative to the oocyte radius), where the upper and lower bounds arise from the slightly ellipsoidal, instead of spherical, shape of the biological spindle. 

While the highly localised nature of the slip velocity in Eq.~\eqref{eq:U_theta_slip_velocity} is inspired by Fig.~\ref{fig:diagram_motivating_spindle_stability}E (reproduced from~\cite{yi2011dynamic}), we note that the experimental mean tangential cortical flow in Fig.~\ref{fig:plot_Niwayama_2016_least_squares_nlinfit_plot_v4_27022024_with_inset_v1_27022024}A (data from~\cite{niwayama2016bayesian}) decreases to zero more gradually as $\theta$ increases to $\pi$, instead of decreasing to zero past a threshold angle~$\theta_\text{slip}$ as in our model. 
Additional simulations using an alternative expression for the slip velocity $U(\theta)$ that includes this long tail yield similar results to those for Eq.~\eqref{eq:U_theta_slip_velocity}; this suggests that the stability properties found in this paper are generic and not sensitive to the precise form of the cortical boundary condition driving the cytoplasmic flow far from the apex of the cell.

\subsection{Analytical model for spindle dynamics}\label{sec:simple_physical_model}

Here, we use lubrication theory to solve the simple analytical model introduced in Sec.~\ref{sec:force_simple} for the fluid flow, pressure and hence hydrodynamic force on the spindle. 
We then conduct a linear stability analysis of the spindle in the flow.

\subsubsection{Setup for analytical model}\label{sec:simple_physical_model_setup}

We illustrate the setup for our analytical model in Fig.~\ref{fig:fig_setup_fixed_spindle_near_cap_wall_lubrication_theory_v7_19042024_with_BC_v4_06052024_plot_28_64_spindle_fixed_circle_wall_lub_theory_plot_force_v7_23042024_with_inset_v1_23042024_combined_v3_13062024}A (see also Sec.~\ref{sec:force_simple}). 
In Cartesian coordinates $(x,y)$, the centre of the spindle has coordinates $(x_\text{sp}, -(d+a))$, where we denote the clearance by $d$ and the horizontal perturbation in spindle position by $x_\text{sp}$. 
As in our numerical simulations (Sec.~\ref{sec:forces_spherical} and Sec.~\ref{sec:spindle_model}, we assume that the clearance $d$ is a constant, with hydrodynamic attraction of the spindle to the cortex balanced by non-hydrodynamic repulsion from the cortex.
On the spindle, which is held stationary as in Sec.~\ref{sec:spindle_model}, we prescribe a no-slip boundary condition given by $\mathbf{u} = \mathbf{0}$, where $\mathbf{u}\equiv(u,v)$ is the fluid velocity field.

On the planar wall, we prescribe a slip velocity boundary condition, given by
\begin{align}
	(u,v) = (U(x),0)  \text{ on $y=0$}.\label{eq:lub_slip_velocity_BC}
\end{align}
In our simple analytical model, the active slip velocity $U(x)$ is given by
\begin{align}
	U(x) = 
	\begin{cases}
		0, & x < -\ell,\\
		-U_\text{max} \sin^2\left (\frac{\pi x}{\ell}\right ), & -\ell \leq x < 0,\\
		U_\text{max} \sin^2\left (\frac{\pi x}{\ell}\right ), & 0 \leq x \leq \ell,\\
		0, & x > \ell,
	\end{cases}
	\label{eq:U_lub_slip_velocity}
\end{align}
where $U_\text{max}$ is the maximum slip speed and $\ell$ is the size of the active slip domain (analogous to the slip angle $\theta_\text{slip}$ in the spherical model).

In all plots in Sec.~\ref{sec:force_simple} and Sec.~\ref{sec:phase_diagram_simple}, we nondimensionalise lengths by $10d$, so that the dimensionless clearance is $d=0.1$, the same numerical value as for our earlier model~(Sec.~\ref{sec:spindle_model}); we also nondimensionalise velocity by the maximum slip velocity $U_\text{max}$ and viscosity by $\mu$. 

\subsubsection{Governing equations}\label{sec:governing_simple}

From biological measurements (e.g.~Fig.~\ref{fig:diagram_motivating_spindle_stability}B), we observe that the clearance $d$ between the spindle and cell cortex is much smaller than the size of the spindle. 
Motivated by this, we may employ the lubrication (or long-wavelength) approximation~\cite{kimbook}, where the characteristic vertical length scale, given by the clearance, is much smaller than the characteristic horizontal length scales, which are set by the geometry and active slip domain size. 
Mathematically, we require $d \ll a, \ell$. 
This is a classical approximation in microhydrodynamics, which allows us to make analytical progress. 
We also assume that the horizontal perturbation in spindle position $x_\text{sp}$ is small, in order to examine the stability of the fixed point at $x_\text{sp}=0$, while neglecting the curvature of the cell cortex. 
Under these assumptions, the dominant contribution to the force on the spindle comes from the thin film of fluid between the spindle and cell cortex, where the fluid flow is mostly horizontal. 
This  model captures the local hydrodynamic effects at play in our earlier, more detailed spherical model solved numerically.

In the lubrication limit, the momentum equations in two dimensions become
\begin{align}
	\frac{\partial p}{\partial x} &= \mu \frac{\partial^2 u}{\partial y^2},\\
	\frac{\partial p}{\partial y} &= 0,
\end{align}
where we recall that $u$ is the $x$ component of the fluid velocity field, 
$p$ is the dynamic pressure and $\mu$ is the dynamic viscosity of the fluid.
The fluid flow is incompressible, which is written as
\begin{align}
	\nabla \cdot \mathbf{u} = 0,
\end{align}
where $\mathbf{u} $ is the fluid velocity field.

We write the equation for the upper boundary of the spindle (close to the cell cortex) as
\begin{align}
	y&=-h(x) \nonumber\\
	&\simeq -d \left (1+\frac{(x-x_\text{sp})^2}{2ad} \right ),\label{eq:h_lub}
\end{align}
where in the second line we have made a local approximation to the circle as a parabola, the standard way of treating this geometry in lubrication theory.
On the spindle, which is stationary and rigid, we prescribe a no-slip boundary condition, given by
\begin{align}
	\mathbf{u}=\mathbf{0} \text{ on $y=-h(x)$}.
\end{align}
The active slip velocity on the planar, active wall $y=0$, which drives the fluid flow, is given by Eqs.~\eqref{eq:lub_slip_velocity_BC}--\eqref{eq:U_lub_slip_velocity}.

\subsubsection{Solution for flow}

Integrating the lubrication equations gives the horizontal fluid velocity as
\begin{align}
	u = \frac{1}{2\mu} \frac{\partial p}{\partial x} (y^2 + hy) + U \left (1+\frac{y}{h}\right ). \label{eq:u_lub}
\end{align}
We integrate this again to find the depth-integrated flux $q$ as
\begin{align}
	q &\equiv \int_{-h}^0 u \, dy \nonumber\\
	&= -\frac{h^3}{12\mu} \frac{\partial p}{\partial x} + \frac{1}{2} U h.
\end{align}
This is a constant (i.e.~independent of $x$), by mass conservation.
We rearrange this for the pressure gradient as
\begin{align}
	\frac{\partial p}{\partial x} = \frac{6\mu}{h^3}  (Uh - 2q ),
\end{align}
which we integrate to give the pressure field as 
\begin{align}
	p(x) &=6 \mu \int_{-\infty}^x \left ( \frac{U(x')}{h(x')^2} - \frac{2q}{h(x')^3} \right ) \, dx'.
\end{align}
The lower limit of integration ensures that the pressure tends to zero as $x \to -\infty$.
Imposing also that $p\to 0 $ as $x \to \infty$, we find the flux $q$ as
\begin{align}
	q = \frac{\int_{-\infty}^\infty \frac{U(x')}{h(x')^2} \, dx'}{2 \int_{-\infty}^\infty \frac{1}{h(x')^3} \, dx'}.
\end{align}

We now write down the pressure gradient in terms of known quantities only as
\begin{align}
	\frac{\partial p}{\partial x} 
	&= 6\mu \left (\frac{U(x)}{h(x)^2} - \frac{\int_{-\infty}^\infty \frac{U(x')}{h(x')^2} \, dx'}{h(x)^3 \int_{-\infty}^\infty \frac{1}{h(x')^3} \, dx'} \right ).
\end{align}
The horizontal velocity is thus given by
\begin{align}
	u(x,y) = 3 \left (\frac{U(x)}{h(x)^2} - \frac{\int_{-\infty}^\infty \frac{U(x')}{h(x')^2} \, dx'}{h(x)^3 \int_{-\infty}^\infty \frac{1}{h(x')^3} \, dx'} \right ) (y^2 + h(x)y) + U(x) \left (1+\frac{y}{h(x)}\right ),
\end{align}
and the pressure field is given by
\begin{align}
	p(x) 
	&= 6\mu \left ( \int_{-\infty}^x \frac{U(x')}{h(x')^2} \, dx' - \frac{\int_{-\infty}^\infty \frac{U(x')}{h(x')^2} \, dx'}{ \int_{-\infty}^\infty \frac{1}{h(x')^3} \, dx'} \int_{-\infty}^x \frac{1}{h(x')^3} \, dx' \right ).
\end{align}

\subsubsection{Vertical force on spindle}\label{sec:vertical_force}

The force on the spindle is dominated by the pressure contribution from the thin-film region, by classical scaling arguments.  
We write down the vertical force on the spindle $F_y$ as
\begin{align}
	F_y = -\int_{-\infty}^\infty p \, dx. \label{eq:Fy}
\end{align}
Here, we have applied the divergence theorem and force balance in Stokes flow to transfer the integral over the upper surface of the spindle $y=-h(x)$ to the wall $y=0$.

We now consider the sign of the vertical force $F_y$.
We note that the vertical force may also be written, by integration by parts, as
\begin{align}
	F_y = \int_{-\infty}^\infty x \frac{\partial p}{\partial x}  \, dx. 
\end{align}
We first consider symmetric positioning of the spindle, with $x_\text{sp}=0$.
In this case, the term $\int_{-\infty}^\infty \frac{U}{h^2} \, dx$ is the integral of an odd function (as $U$ is odd and $h$ is even) over a symmetric range, and is hence equal to zero.
The vertical force on the spindle is therefore given by
\begin{align}
	F_y \vert_{x_\text{sp}=0} 
	= 6\mu \int_{-\infty}^\infty  \frac{x U}{h^2}  \, dx.
\end{align}
This is positive, since the integrand is positive.
Therefore, by continuity, if the spindle is sufficiently close to $x_\text{sp}=0$, then $F_y$ is still positive.
In Sec.~\ref{sec:linear_stability}, we will consider the linear stability of the fixed point $x_\text{sp}=0$. 
Since the vertical force $F_y$ always pulls the spindle towards the cell cortex for small $x_\text{sp}$,  the sign of the horizontal force $F_x$ solely determines the linear stability of the fixed point.

\subsubsection{Horizontal force on spindle}\label{sec:horizontal_force}

In the lubrication limit, the horizontal force on the spindle $F_x$ is given by 
\begin{align}
	F_x &= \int_{-\infty}^\infty \left ( -p \frac{dh}{dx} + \mu \left .\frac{\partial u}{\partial y} \right \vert_{y=-h} \right ) \, dx ,\label{eq:Fx_suction_shear}
\end{align}
and is the sum of a horizontal suction force (first term) and a shear force (second term).
We use this decomposition to interpret our stability results in Sec.~\ref{sec:physical_mechanism}. 
By the divergence theorem and force balance in Stokes flow, we may alternatively compute $F_x$ as
\begin{align}
	F_x &= \mu\int_{-\infty}^\infty \left .\frac{\partial u}{\partial y} \right \vert_{y=0}  \, dx.
\end{align}
We finally obtain the horizontal force on the spindle $F_x$ as
\begin{align}
	F_x 
	&= \mu \left ( 4\int_{-\infty}^\infty \frac{U}{h} \, dx - \frac{3 \int_{-\infty}^\infty \frac{1}{h^2} \, dx \int_{-\infty}^\infty \frac{U}{h^2} \, dx}{\int_{-\infty}^\infty \frac{1}{h^3} \, dx} \right ).\label{eq:Fx}
\end{align}

\subsubsection{Linear stability}\label{sec:linear_stability}

To explain the stability results in  Fig.~\ref{fig:plot_1_stability_spindle_fixed_circle_wall_lub_theory_phase_w_bdry_v4_27102023} mathematically, we conduct a linear stability analysis of the fixed point $x_\text{sp}=0$. 
We consider small perturbations of the horizontal position of the spindle $x_\text{sp}$ from the fixed point $x_\text{sp}=0$. 
Recall from Sec.~\ref{sec:vertical_force} that for sufficiently small~$x_\text{sp}$, the vertical component of force on the spindle $F_y$ [Eq.~\eqref{eq:Fy}] is positive, keeping the spindle near the cortex. 
Thus, the sign of the horizontal  force on the spindle $F_x$ [Eq.~\eqref{eq:Fx}] determines the stability of the fixed point. 
We note that for $x_\text{sp}=0$, the horizontal force on the spindle is exactly zero, by symmetry. 
For small $x_\text{sp}$, by changing variables to work in the frame of the spindle and Taylor-expanding Eq.~\eqref{eq:U_lub_slip_velocity}, we obtain the horizontal force $F_x$ to leading order in $x_\text{sp}$ as
\begin{align}
	F_x \simeq& \frac{4\pi \mu U_\text{max} x_\text{sp}}{\ell} \left ( 
	4\int_{0}^{\ell} 
	\frac{
		\sin \left (\frac{\pi x}{\ell}\right ) \cos \left (\frac{\pi x}{\ell}\right )   }{h\vert_{x_\text{sp}=0}} \, dx
	-\frac{3 \int_{-\infty}^\infty \frac{1}{h^2\vert_{x_\text{sp}=0}} \, dx }{\int_{-\infty}^\infty \frac{1}{h^3\vert_{x_\text{sp}=0}} \, dx} \int_{0}^{\ell} 
	\frac{
		\sin \left (\frac{\pi x}{\ell}\right ) \cos \left (\frac{\pi x}{\ell}\right )   }{h ^2\vert_{x_\text{sp}=0}} \, dx \right ) .\label{eq:Fx_expand_1}
\end{align}
For the parabolic height profile of the spindle [Eq.~\eqref{eq:h_lub}], we may evaluate some of the integrals in Eq.~\eqref{eq:Fx_expand_1} analytically to find a simpler expression for $F_x$ as 
\begin{align}
	F_x \simeq  \frac{16\pi \mu U_\text{max} x_\text{sp}}{\ell} \left ( 
	\int_{0}^{\ell} 
	\frac{
		\sin \left (\frac{\pi x}{\ell}\right ) \cos \left (\frac{\pi x}{\ell}\right )   }{h\vert_{x_\text{sp}=0}} \, dx
	-d \int_{0}^{\ell} 
	\frac{
		\sin \left (\frac{\pi x}{\ell}\right ) \cos \left (\frac{\pi x}{\ell}\right )   }{h ^2\vert_{x_\text{sp}=0}} \, dx 
	\right ).
\end{align}
By setting $F_x$ to zero, we obtain the boundary between a linearly stable and unstable fixed point at $x_\text{sp}=0$ as
\begin{align}
	\int_{0}^{\ell} 
	\frac{
		\sin \left (\frac{\pi x}{\ell}\right ) \cos \left (\frac{\pi x}{\ell}\right )   }{h\vert_{x_\text{sp}=0}} \, dx
	-d \int_{0}^{\ell} 
	\frac{
		\sin \left (\frac{\pi x}{\ell}\right ) \cos \left (\frac{\pi x}{\ell}\right )   }{h ^2\vert_{x_\text{sp}=0}} \, dx = 0.\label{eq:stability_boundary_1}
\end{align}
This is an equation in terms of the spindle radius~$a$, the slip domain size~$\ell$ and the clearance~$d$. 
The clearance is a fixed parameter, so the stability boundary in Eq.~\eqref{eq:stability_boundary_1} is a curve in the $(a,\ell)$ plane.
To learn more about the shape of this curve, we use the substitution $s=x/\ell$ in Eq.~\eqref{eq:stability_boundary_1} to rewrite the equation as
\begin{align}
	\int_{0}^{1} 
	\frac{
		\sin (\pi s) \cos (\pi s)   }{ 1 + \frac{1}{2}C^2 s^2 } \, ds
	- \int_{0}^{1} 
	\frac{
		\sin (\pi s) \cos (\pi s)  }{ \left ( 1 + \frac{1}{2} C^2 s^2 \right )^2 } \, ds = 0,\label{eq:stability_boundary_2}
\end{align}
where we define the dimensionless number $C$  as
\begin{align}
	C \equiv \frac{\ell}{a^{1/2} d^{1/2}}.
\end{align}
The stability boundary is therefore simply given by
\begin{align}
	\ell = C d^{1/2} a^{1/2},\label{eq:stability_boundary_3}
\end{align}
where the dimensionless constant $C \approx 3.50$ is found by solving Eq.~\eqref{eq:stability_boundary_2} numerically. 
Hence, the size of the slip domain $\ell$ at the boundary between linear stability and instability of the fixed point scales with the square root of the spindle radius ($\ell \sim a^{1/2}$); the horizontal length scale in Eq.~\eqref{eq:stability_boundary_3} originates from the circular shape of the model spindle in the lubrication limit.
 
\section*{Acknowledgements}
We gratefully acknowledge funding from the Engineering and Physical Sciences Research Council (studentship to W.L.) and Trinity College, Cambridge (Rouse Ball and Eddington Research Funds travel grant to W.L.). 

\bibliography{cytoplasmic_streaming_bibliography_22072024.bib}
 
\end{document}